\documentclass[aps,pra,twocolumn,showpacs,preprintnumbers,amsmath,footinbib,10pt,longbibliography]{revtex4-2}

\pdfoutput=1
\makeatletter
\renewcommand*{\@fnsymbol}[1]{\ensuremath{\ifcase#1\or \or \spin\or \*\or \quadrupole\or
   \mathsection\or \mathparagraph\or \|\or **\or \dagger\dagger
   \or \ddagger\ddagger \else\@ctrerr\fi}}
\makeatother

\usepackage{hyperref}
\usepackage[utf8]{inputenc}
\usepackage[sort&compress]{natbib}
\usepackage[dvipsnames]{xcolor}
\usepackage{graphicx}   
\usepackage{epsfig}
\usepackage{epstopdf}
\usepackage{amsfonts}
\usepackage{amstext,amsmath,amssymb}
\usepackage{svrsymbols}
\usepackage{rotating}
\usepackage{dcolumn}
\usepackage{array,multirow}
\usepackage[english]{babel}
\usepackage[separate-uncertainty=true]{siunitx}

\DeclareSIUnit\stokes{St}
\DeclareSIUnit\gauss{G}
\usepackage[export]{adjustbox}
\usepackage{braket}
\usepackage[caption=false]{subfig}
\usepackage{xr}
\usepackage{ifpdf}
\usepackage{tikz}
\usetikzlibrary{arrows}

\def\ket#1{| #1 \rangle}

\def\avg#1{\left\langle #1 \right\rangle}

\emergencystretch=\maxdimen
\usepackage[none]{hyphenat}

\hypersetup{urlcolor=violet, citecolor=violet, linkcolor=violet, colorlinks=true}
\newcommand{\be}{\begin{equation}}
\newcommand{\ee}{\end{equation}}
\newcommand{\bea}{\begin{eqnarray}}
\newcommand{\eea}{\end{eqnarray}}

\usepackage{color}

\def\brms{B_\text{rms}}
\def\frms{\Phi_\text{rms}}

\begin{document}

\author{Nicolas Staudenmaier${}^{1,\spin,\quadrupole}$, Anjusha Vijayakumar-Sreeja${}^{1,\spin}$, Santiago Oviedo-Casado${}^{2,\spin}$, Genko Genov${}^{1,\spin}$, Daniel Cohen${}^{2}$, Daniel Dulog${}^{3}$, Thomas Unden${}^{3}$, Nico Striegler${}^{3}$, Alastair Marshall${}^{3}$, Jochen Scheuer${}^{3}$, Christoph Findler${}^{1,4}$, Johannes Lang${}^{1,4}$, Ilai Schwartz${}^{3}$, Philipp Neumann${}^{3}$, Alex Retzker${}^{2,5}$, Fedor Jelezko${}^{1}$\vspace{5pt}}

\affiliation{${}^{1}$Institute for Quantum Optics, Ulm University, Albert-Einstein-Allee 11, 89081 Ulm, Germany\\
${}^{2}$Racah Institute of Physics, The Hebrew University of Jerusalem, 91904 Givat Ram, Jerusalem, Israel.\\
${}^{3}$NVision Imaging Technologies GmbH, Albert-Einstein-Allee 11, 89081 Ulm, Germany\\
${}^{4}$Diatope GmbH, Buchenweg 23, 88444 Ummendorf, Germany\\
${}^{5}$AWS Center for Quantum Computing, Pasadena, CA}

\email{${}^{\quadrupole}$nicolas.staudenmaier@uni-ulm.de}
\thanks{\\${}^{\spin}$These authors contributed equally}

\title{Power-law scaling of correlations in statistically polarised nano-NMR}

%\date{\today}

\begin{abstract}
Diffusion noise is a major source of spectral line broadening in liquid state nano-scale nuclear magnetic resonance with shallow nitrogen-vacancy centres, whose main consequence is a limited spectral resolution. This limitation arises by virtue of the widely accepted assumption that nuclear spin signal correlations decay exponentially in nano-NMR. However, a more accurate analysis of diffusion shows that correlations survive for a longer time due to a power-law scaling, yielding the possibility for improved resolution and altering our understanding of diffusion at the nano-scale. 
Nevertheless, such behaviour remains to be demonstrated in experiments. 
Using three different experimental setups and disparate measurement techniques, we present overwhelming evidence of power-law decay of correlations. These result in sharp-peaked spectral lines, for which diffusion broadening need not be a limitation to resolution. 
\end{abstract}

\maketitle

%%%%%%%%%%%%%%%%%%%%%%%%%%%%%%%%%%%%%%%%%%%%%%%%%%%%%%%%%%%%%%%%%%%%%%%%%%%%%%%%%%%%%%%%%%%%%%%%%%%%%%%%%%%%%%%%%%%%%%%%%%%%%%%
%%%%%%%%%%%%%%%%%%%%%%%%%%%%%%%%%%%%%%%%%%%%%%%%%%%%%%%%%%%%%%%%%%%%%%%%%%%%%%%%%%%%%%%%%%%%%%%%%%%%%%%%%%%%%%%%%%%%%%%%%%%%%%%
\section{Introduction}\label{Section:Introduction}
%%%%%%%%%%%%%%%%%%%%%%%%%%%%%%%%%%%%%%%%%%%%%%%%%%%%%%%%%%%%%%%%%%%%%%%%%%%%%%%%%%%%%%%%%%%%%%%%%%%%%%%%%%%%%%%%%%%%%%%%%%%%%%%
%%%%%%%%%%%%%%%%%%%%%%%%%%%%%%%%%%%%%%%%%%%%%%%%%%%%%%%%%%%%%%%%%%%%%%%%%%%%%%%%%%%%%%%%%%%%%%%%%%%%%%%%%%%%%%%%%%%%%%%%%%%%%%%

%All throughout the text we have to check for consistency on power-law decay, polynomial, correlations, correlations decay, and so on. 

Nuclear magnetic resonance (NMR) spectroscopy is widely used in the life and material sciences. However, classical NMR techniques are not readily available on the single cell level. Therefore, further scientific breakthrough in biosensing may be hindered by the invasive analysis techniques that most conventional approaches require in this regime. Existing methodologies entail tagging, e.g., with fluorescent nano-particles, cryogenic temperatures, or high magnetic fields \cite{Chen2006,Eggeling2009,Kovacs2005,Aslam2017,Bucher2020}. Either causes substantial modifications to the sample, meaning that its natural properties cannot be determined accurately. Nuclear magnetic resonance at the nano-scale (nano-NMR) with spin sensors, such as nitrogen-vacancy (NV) centres, offers a label-free, room temperature approach capable of studying biologically relevant samples down to the molecular level without altering their properties \cite{Cappellaro2017,Degen2017,Ajoi2015,Lovchinsky2016}. NV centres have already been shown capable of detecting the magnetic field created by nano-sized distributions of nuclei in liquid samples at ambient temperature \cite{Staudacher2013,Glenn2018,bucher2020hyperpolarization,devience2015nanoscale,loretz2014nanoscale,Aslam2017}. However, a fundamental limitation to the ability of resolving spectral lines is thought to exist. 

\begin{figure}[t!]
\includegraphics[width=\columnwidth]{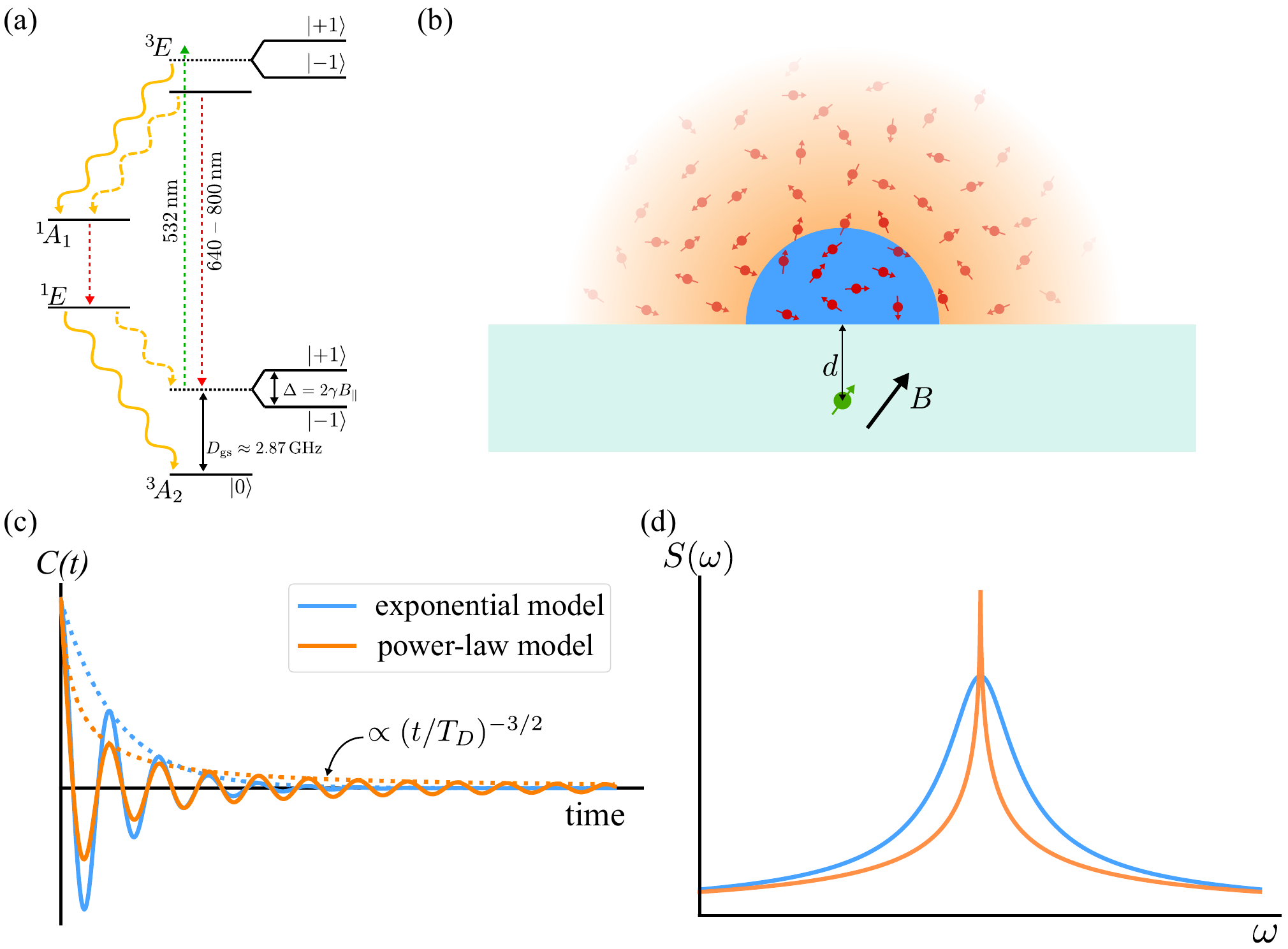}
\caption{NV level scheme and power-law model. (a) Level scheme of the NV centre. (b) The NV centre is located at a depth $d$ below the diamond surface. In the standard diffusion model a hemisphere sensing volume (blue) of radius $\sim\!d$ is considered, in which the nuclei interact with the NV centre. Alternatively, we accurately take into account the $1/r^3$ dependence of the dipole-dipole interaction. The colour intensity of the orange hemisphere indicates the interaction strength of the nuclear spins (red) with the NV centre. (c) Decay of the auto-correlation according to the exponential and the power-law model. With the accurate dipole-dipole model, decay is (nearly) exponential for short times ($t\ll T_D$) and a power-law for $t\gg T_D$ with $T_D$ the characteristic diffusion time. (d) Power spectrum of the NMR signal. The power-law model shows a distinct, sharp peak whereas the exponential model resembles a Lorentzian lineshape.}
\label{Fig:diffusion_scheme}
\end{figure}

The nano-NMR approach with spin sensors relies on nano-scale sample sizes containing a sufficiently low number of nuclei, in which statistical fluctuations are significant enough to overcome the thermal averaging of nuclei orientation, leading to the emergence of time-correlated magnetic fields without resorting to sample polarisation. These fields can be detected at room temperature with quantum probes such as a shallow NV centre, whose interaction region is typically considered a hemisphere with a radius of the order of the NV depth as shown in Fig.~\ref{Fig:diffusion_scheme}(b) \cite{Wrachtrup2015,Pham2016,Aslam2017,Fernandez2017}. Sample sizes within the NV detection region are on the order of $10^{-24}\!-\!10^{-20}$\,L for several nm deep NV centres. There, the statistical polarisation exceeds the thermal by more than three orders of magnitude in ambient conditions \cite{Degen2007,Reinhard2012,Staudacher2013,Herzog2014,Mamin2013,Muller2014}. Statistically polarised nano-NMR with spin sensors enables studying sample properties inaccessible with any other NMR protocol.

The main challenge for nano-NMR with statistically polarised liquid samples is posed by the diffusion of molecules out of the interaction region defined by the probe. Diffusion changes the spatial distribution of statistical polarisation, and leads to a decay of the correlations of the magnetic field signal in the probe, which is typically considered exponential (see Fig.~\ref{Fig:diffusion_scheme}(c)) \cite{HubbardPR1963,Pham2016}. Then, measurements performed beyond the characteristic time of the exponential decay yield no information about the spectral signatures of interest. When this time is short, not enough information can be gathered to allow for spectral reconstruction in post-processing \cite{Oviedo2020}, creating a resolution problem. The corresponding spectral line for correlations with exponential decay is Lorentzian, as shown in Fig.~\ref{Fig:diffusion_scheme}(d), where frequency resolution is typically defined by its full width at half maximum \cite{Abbe1873,Rayleigh1879,JonesAR1995}. 

Recent theoretical analysis has claimed that diffusion induced decay of correlations follows a power-law at long times (see Fig.~\ref{Fig:diffusion_scheme}(c)), due to the specific dipole-dipole interaction of the shallow NV centre sensors and the nuclei in the sample \cite{Cohen2020}. Such correlations allow suitable measurement protocols to extract sufficient information beyond the characteristic decay time $T_D$, so estimating frequencies smaller than $1/T_D$ becomes feasible \cite{Oviedo2020}. Moreover, they provide a more accurate route to measure the diffusion coefficient in liquid samples. Although previous work in liquid NV nano-NMR has shown an instance where a non-exponential function fits better the observed correlation, this was attributed to surface effects that led to reduced translational diffusion of the nuclei close to the surface \cite{Wrachtrup2015}. Thus, experimental evidence of interaction dependent power-law behaviour has not been demonstrated yet, to the best of our knowledge. The main reason might be that the early time decay of correlations resembles an exponential function, and it is only the long-time decay beyond the diffusion time that is critical to show deviations from the exponential paradigm \cite{Farida2021}.

In this work, we provide compelling experimental evidence of a power-law decay of correlations at long times, in accordance with the theoretical prediction in Ref. \cite{Cohen2020}. We demonstrate our results using three distinct measurement settings and several independent statistical \mbox{analysis} tools. We include measurements of the magnetic field originated in the fluorine nuclear spins of the sample in one of the experiments. These nuclei are assumed to be separated from the diamond surface by an immobilised proton layer adsorbed onto the surface, and permit us to rule out surface effects as the source of a long correlation tail \cite{Wrachtrup2015}. Our results pave the way for statistically polarised, room temperature nano-NMR experiments with biomedically relevant samples, with a scope that is not limited to NV centres but that, due to the power-law scaling stemming from dipole-dipole coupling, extends to any sensor that is based on such interactions.

\section{Results}

\subsection{Theory}

The NV centre is a point defect in the diamond lattice composed of a substitutional nitrogen atom and an adjacent vacancy which, together, are described as a system whose ground state is a $^3A_2$ spin triplet. Applying a bias magnetic field, the degeneracy of the $\ket{m_s=\pm 1}$ spin levels is lifted, and the NV centre can be used as an effective two-level system, as shown in Fig.~\ref{Fig:diffusion_scheme}(a). Owing to its spin properties, an initial equal superposition state of two of the relevant states (e.g. $\{\ket{0},\ket{-1}\}$) accumulates a phase due to interaction with an external magnetic field. A rotation of the evolved state into the measurement basis reflects the accumulated phase as a population imbalance between the NV centre spin states, which can be measured thanks to the distinct fluorescence rates of each of the states. Such an initial state is most sensitive to the slowest frequency components of the (magnetic) noise. Therefore, dynamical decoupling (DD) sequences are routinely used to filter out the effect of slow noise, prolong coherence times, and enable sensing of specific frequency components of the signal \cite{Viola1999,Cywinski2007,Degen2007}. %They rotate the frame of reference of the NV close to the frequency of the signal to be detected, which contributes most to dephasing and detection.

In statistically polarised nano-NMR, the magnetic signal originates on an ensemble of nuclei within a small sensing volume oscillating at their Larmor frequency. Diffusion of molecules induces magnetic noise which we model as a stationary Gaussian process with zero mean. For all experiments, we estimate the possible spectral broadening effect from back-action on the NV centre or a gradient field due to the NV sensor \cite{Unden2018} to be at least one order of magnitude smaller than the broadening caused by diffusion (see Appendix~\ref{Section:Mixed noise model} for details). Therefore we consider it negligible for the ensuing analysis of the decay of correlations due to diffusion. Throughout the experiments described here, we will be probing the auto-correlation of the time evolution of the NV centre interacting with the nuclear spins, which is given by the noise covariance of the external signal,
\begin{equation} 
\Phi_{\text{rms}}^2\cos(\delta t)C(t/T_D). 
\label{basicmodel}
\end{equation} 
Here, $\Phi_{\text{rms}}$ is the accumulated phase on the NV probe, which is proportional to the root-mean-square field $\brms$ originating from the statistically polarised nuclei, 
and $\delta$ is the typically small undersampling frequency of the detected signal. The latter is equal to the difference between the Larmor frequency and the frequency defined by the sampling times in the measurement protocol. The envelope $C(t/T_D)$ describes the effect of noisy fluctuations due to diffusion, with characteristic diffusion time $T_D=\frac{d^2}{D}$ where $D$ is the diffusion coefficient \cite{Pham2016}. Its particular shape is determined by the specific interaction between the sample and the sensor. For NV centres, it is the magnetic dipole-dipole interaction with the nuclei. In the common paradigm, this interaction lasts for $T_D$. Thus, nuclei that diffuse beyond a distance $\sqrt{T_DD} \sim d$ are too far to interact with the NV centre. This defines an interaction volume which is a hemisphere of radius $\sim d$ above the surface of the diamond, illustrated by the solid colour region in Fig.~\ref{Fig:diffusion_scheme}(b) \cite{Pham2016}. Correlations of the magnetic field at the NV position reflect the diffusion of nuclei in or out of this hemisphere, and the assumption about the interaction duration means that correlations are lost when nuclei diffuse out of the interaction region. In this picture $C(t/T_D) \propto \exp(-t/T_D)$. The exponential model is widely accepted in the field due to its success when the measurement times are short and the frequencies probed are easily resolved \cite{Jelezko2015, Wrachtrup2015, Pham2016}. However, such approximation has profound implications over measurements about the diffusion coefficient $D$ in microfluids \cite{Cohen2020,Farida2021}. Moreover, it results in a resolution problem which cannot be overcome \cite{Oviedo2020}. 

A more careful analysis of the dipole-dipole interaction renders a significantly different behaviour \cite{Cohen2020}. Heuristically, it can be understood as follows: the magnetic field generated at the NV position as a result of a nucleus located at a point $\vec{r}$ is $B(t)\propto \frac{1}{r^3}$. Hence, the correlation of the magnetic field is $\left<B(t)B(0)\right>\propto\left<\frac{1}{r^3}\frac{1}{r'^3}\right>$, where the coordinates $\vec{r}$ and $\vec{r}'$ of a specific nucleus are related by its diffusive motion. This connection is usually observed in the second moment 
$\left<r'^2\right>=\left<x'^2\right>+\left<y'^2\right>+\left<z'^2\right>=x^2+y^2+4Dt+\left<z'^2\right>\approx r^2+6Dt$, where $2Dt$ is the variance in the nuclei positions in one dimension. Diffusion of molecules close to the surface is free along the surface direction $x$ and $y$ but limited in the orthogonal $z$. Hence, the second equality is always valid, while the last approximation is correct for $t\ll T_D$ or $t\gg T_D$, where the exact geometry of the problem is less important. Substituting the second moment into the auto-correlation function yields $\left<B(t)B(0)\right>\propto\left<\frac{1}{r^3}\frac{1}{\left(r^2+6Dt\right)^{3/2}}\right>$. Averaging over the effective interaction volume leads to the approximate form $\left<B(t)B(0)\right>\propto\frac{1}{d^3}\frac{1}{\left(d^2+6Dt\right)^{3/2}}$. At short times $t\ll T_D$ only close nuclei contribute, the hemisphere region approximation is valid, and correlations replicate the exponential behaviour. At long times, the interaction region expands beyond the hemisphere paradigm, as shown in Fig.~\ref{Fig:CS_NV4_v3}(b), and correlations decay as a power law $\left<B(t\gg T_D)B(0)\right> \propto \frac{1}{\left(6Dt\right)^{3/2}}$. The full expression for $C(t/T_D)$ in this scenario is given by $G(t/T_D)$ in Eq.~\eqref{Cohenian} (see Methods). A more detailed derivation of the asymptotic behaviour of the power spectrum due to these correlations shown in Fig.~\ref{Fig:diffusion_scheme}(d) can be found in Appendix~\ref{Cohenianderivation} and in Ref.~\cite{Cohen2020}.

\subsection{Experiments}

We performed a series of experiments comprising three different setups: correlation spectroscopy measurements with single NV centres and with an ensemble of NV centres, and quantum heterodyne (Qdyne) with single NV centres. The experiments and results are reported in each subsection, altogether reinforcing the validity of the power-law scaling of the correlation's decay. Description of the measurement sequences can be found in Section \ref{Section:Methods} (Methods). All experimental details and parameters are thoroughly described in Appendix \ref{Section:Experimental_details}.

\subsubsection{Correlation spectroscopy with a single NV centre}

The first set of experiments is carried out on a shallow NV centre which is located at a depth $\approx 2.9$\,nm below the surface of an isotopically enriched (99.999$\%$ $^{12}C$) diamond sample grown by chemical vapour deposition \cite{osterkamp2019engineering,silva2010microwave}.
The shallow NV centre is addressed via a fluorescent confocal scan and is initialised and read out using a 532\,nm laser pulse. We apply a bias field of $\approx 450$\,G along the NV axis using a permanent magnet to lift the $m_s = \pm 1$ degeneracy (see Fig.~\ref{Fig:diffusion_scheme}(a)). Microwave pulses for coherent control of the NV centre are applied through a copper wire of $20\,\mu$m diameter strapped across the diamond, and we use state-dependent photoluminescence measurements to detect the population of the spin states. The interaction of the NV centre with hydrogen nuclear spins in the immersion oil (Fluka 10976, viscosity of 400\,cSt) is measured. 

\begin{figure}[t!]
\includegraphics[width=0.9\columnwidth]{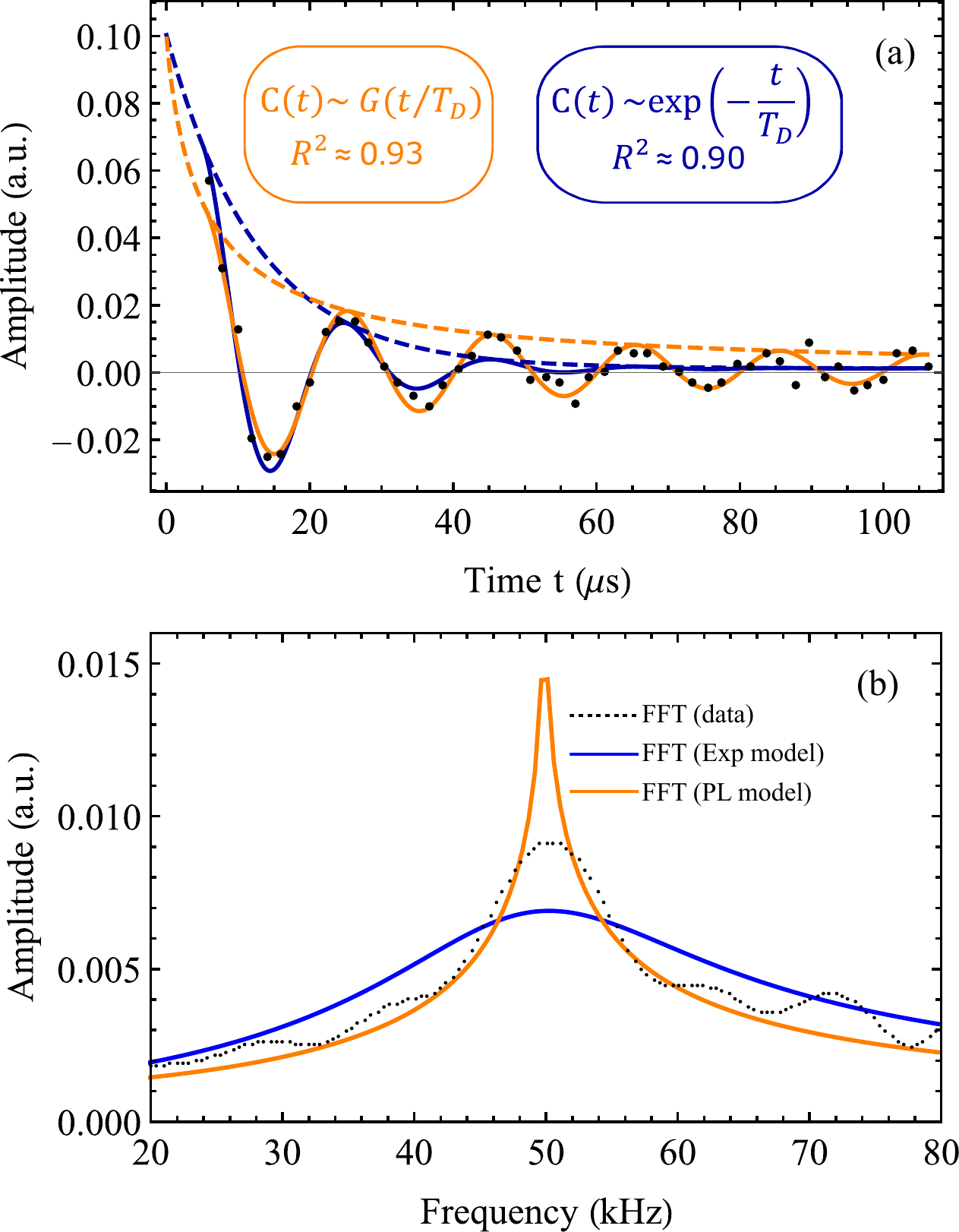}
\caption{Experimental data from correlation spectroscopy measurement with a single NV centre. (a) Signal of the correlation spectroscopy measurement as a function of the time difference between the two DD sequences (black dots) and fits of the exponential (blue line) and power-law (orange line) models. Labels show the goodness of fit. The power-law model, described by Eq.~\eqref{Cohenian}, offers a better fit to the data, which can be appreciated also in the $C(t/T_D)$ envelope (dashed lines). In both models, the initial amplitude for the fitting algorithm is fixed, obtained from an independent estimate in a power spectrum measurement. 
(b) Fast Fourier transform (FFT) of the experimental data together with FFT from the fitted data in (a) to the exponential and the power-law decay models, shown for illustrative purposes. The sharpness displayed by the power law model allows for a more precise frequency estimation than the exponential model.
}
\label{Fig:CS_NV4_v3}
\end{figure}

The NV centre is initialised into its $m_s=0$ state, and read out optically. We use correlation spectroscopy to probe the diffusion spectrum (see Methods \ref{Subsection:CS} for details). First, we prepare the system in state $\ket{Y} = \frac{\ket{0} + i \ket{1}}{\sqrt{2}}$ by applying a $\pi/2$ pulse around $x$-axis and use two Knill dynamical decoupling (KDD4) sequences, separated by a waiting time difference $T$ \cite{KDDref,KDDref2,GenovPRL2017,CasanovaPRA2015}.
The centres of two subsequent $\pi$ pulses in the sequence are separated by the time $\tau=1/2f_\mathrm{L}$, where $f_\mathrm{L}$ is the Larmor frequency of the nuclear spins \cite{KDDref,KDDref2,GenovPRL2017,CasanovaPRA2015}, which allows for maximum phase accumulation of the NV centre due to the interaction with the nuclear spins. In this case, the Larmor frequency is 1.9 MHz. The information about the accumulated phase during the first decoupling sequence is mapped onto the spin population by another $\pi/2$ pulse applied around the $y$-axis. This result is then correlated with the second interrogation starting at time $t=N\tau+T$ of the second KDD4 decoupling sequence, where $N=20$ is the number of pulses in KDD4. %In the end, the protocol is limited by $T_1$ of the NV centre. 
The obtained signal is proportional to the correlation of the accumulated phases during the two sequences, and allows us to directly extract the auto-correlation of the signal.

Figure \ref{Fig:CS_NV4_v3}(a) shows the measured signal vs. the time $t$ between the beginnings of the two DD sequences, fitted to the exponential and the power-law decay models described by the correlation function Eq.~\eqref{basicmodel} with envelopes Eqs.~\eqref{expmodel} or \eqref{Cohenian}, respectively. Note that Eq.~\eqref{Cohenian} comes from exact mathematical calculations such that the envelope is valid for any time and shows a power-law decay for longer correlation times.
We restrict the parameter space for both models by estimating the initial value of the signal amplitude $\sim\Phi_{\text{rms}}^2$ from an independent power spectrum measurement, which is typically used for finding the depth of the NV centre (see \cite{Pham2016} and subsection \ref{Appendix:CS_single_NV} in the Appendix). %
This is preferable, as the signal sampling times are chosen for better estimation of long-lived correlations. Thus, the estimate of the initial amplitude is not efficient and differs significantly between the two models, making comparison difficult. 
We then use the estimated value as a fixed input parameter for both models and apply non-linear least squares fitting of the signal. We start from 500 different initial conditions and take the best fit.

The obtained fits in Fig.~\ref{Fig:CS_NV4_v3}(a) show that the power-law decay model performs better with an $R^2\approx 0.93$ in comparison to $R^2\approx 0.90$ for the exponential model. Its goodness of fit is evident especially at long times, e.g., between $50-100\,\mu$s, which correspond to more than three times the expected diffusion time $T_D\approx 17\,\mu$s, highlighting the importance of long-lived correlations for an accurate estimation of the diffusion coefficient, and emphasising the need to estimate independently the initial contrast and the characteristic decay time in order to obtain accurate measurements of the diffusion coefficient. 

In addition, Fig.~\ref{Fig:CS_NV4_v3}(b) shows a comparison of the Fast Fourier transform (FFT) of the experimental data and the FFT from the fitted data to the exponential and the power-law decay models. The results show that the power-law decay model provides a better fit to the data than the exponential model. Its spectral linewidth is narrower than with the exponential fit, which is due to the long-lived correlations of the power-law model. Such a feature enables improved precision and resolution in sensing experiments due the sharply peaked spectrum of the power-law model (see also Fig.~\ref{Fig:diffusion_scheme}(d)). The slight broadening of the experimental FFT is mainly due to the fitting procedure, which uses zero padding, so we prolong the time domains of the fits of both theoretical models accordingly. This results in a sharper peak for the power-law model, as expected from theory. Such broadening can also be present due to extra noise sources which produce exponential decay at a much slower rate than that of diffusion noise, as described in Appendix~\ref{Section:Mixed noise model}.

\subsubsection{Correlation spectroscopy with an NV centres ensemble}

The next set of experiments is carried out on an ensemble of NV centres. 
Similarly to the single NV experiment, we use correlation spectroscopy to probe the diffusion spectrum. Here, a perfluoropolyether oil (Fomblin Y, Sigma Aldrich 317926, viscosity of 60\,cSt) is analysed. We detect the signal coming from nuclei in the Fomblin oil, which sits above a proton layer located on top of the diamond surface, preventing fluorine nuclei from sticking to the surface. Thus, we can rule out surface effects as the underlying cause of the power-law behaviour.

The correlation spectroscopy protocol is the same as in the single NV experiment, except that each of the two DD sequences is KDD4-4, i.e., KDD4, repeated four times, accounting for the larger average depth of the NV centres in the ensemble of about 9\,nm and the higher Larmor frequency, for which the centres of the decoupling pulses are separated by $\tau=1/2f_\mathrm{L}=135.6$\,ns ($f_\mathrm{L}=3.687$\,MHz).

\begin{figure}[t!]
\includegraphics[width=0.9\columnwidth]{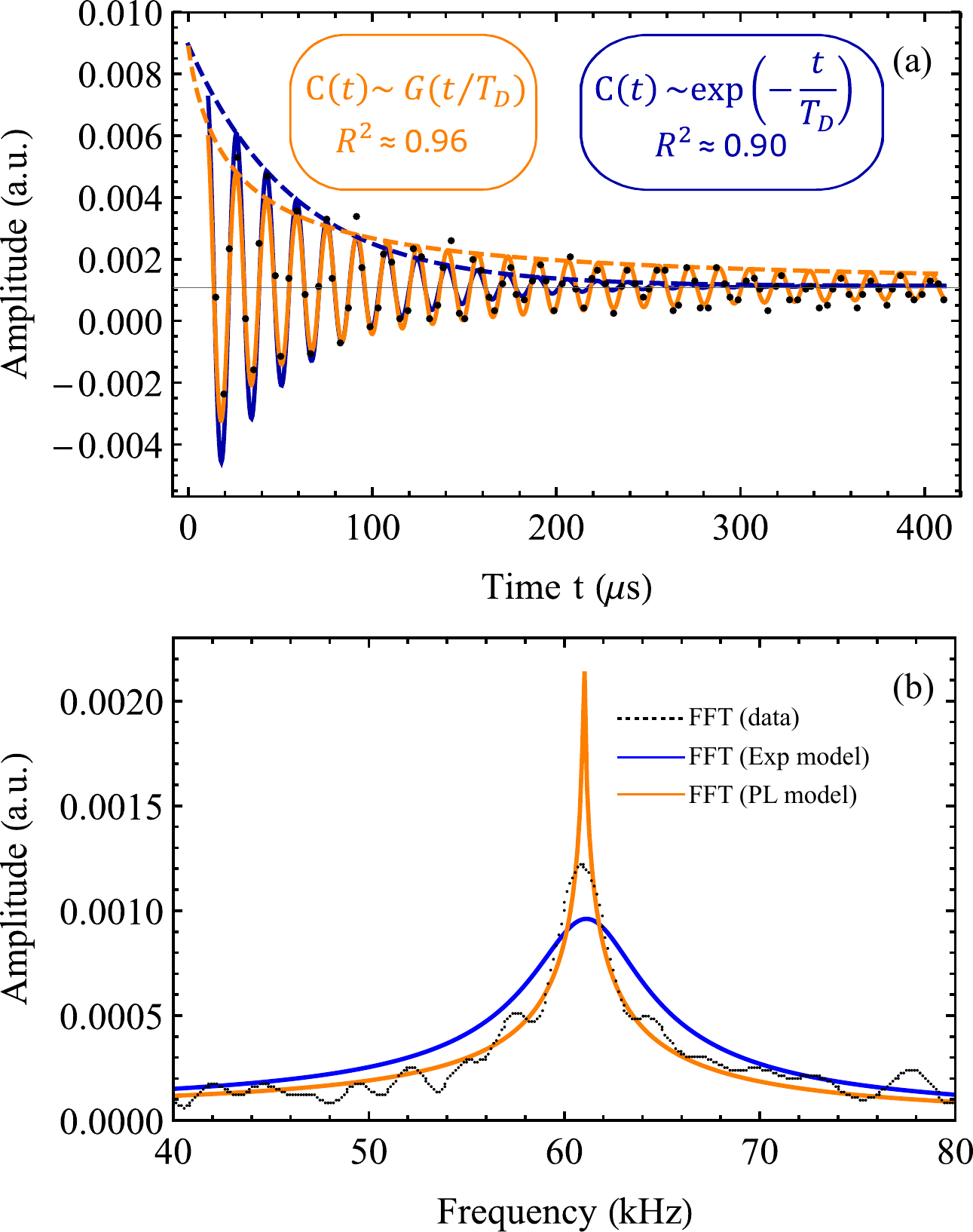}
\caption{Experimental data from correlation spectroscopy measurement with an ensemble of NV centres. (a) Signal of the correlation spectroscopy measurement vs. the time difference between the two DD sequences (black dots) and fits of the exponential (blue line) and power-law models (orange line) with their goodness of fit and the corresponding envelopes shown as dashed lines. The power law model shows a better fit to the data, especially at long times. In both fits we fixed the initial amplitude of the fitted signal to an estimate from an independent power spectrum measurement. % $a_1$. 
%(b) Same as (a) with a fixed initial amplitude of the oscillation, % $a_1=0.12$, estimated from a power spectrum measurement.
(b) Fast Fourier transform (FFT) of the experimental data together with the FFT from the fitted data to the exponential and the power-law decay models, shown here for illustrative purposes. Similarly to the single NV centre experiments, the power law model provides a better fit of the sharp peaked spectral line, thus allowing for better frequency estimation.
}
\label{Fig:CS_NVision}
\end{figure}

Figure \ref{Fig:CS_NVision}(a) displays the measured signal vs. the time $t$ between the beginnings of the two DD sequences, together with its fit to the two decay models, exponential and power-law, as described. %similarly to Fig.~\ref{Fig:CS_NV4_v3}.
We use the same data analysis approach as for the single NV centre experiments. %Again, the results for both models fit the data well with $R^2\approx 0.96$.
Since the measurements focus on targeting the later times of the correlations, sampling times are chosen accordingly. Thus, instead of considering the signal amplitude as a free parameter, which is inefficient and leads to big differences between models (see subsection \ref{Appendix:CS_ensemble_NV} in the Appendix), we estimate it from independent power spectrum measurements and provide it as a fixed input parameter to each of the 500 non-linear least squares fittings of the signal, analogously to the single NV experiment, and consider the best fit.
The obtained fits in Fig.~\ref{Fig:CS_NVision}(a) show that the power-law decay model provides a better fit to the experiment, in line with the results obtained with the single NV centre. The goodness of the fit of the power law model is especially better at long times, e.g., between $200-400\,\mu$s, similarly to the single NV centre experiment. 
Finally, Fig.~\ref{Fig:CS_NVision}(b)  compares the FFT of experimental data and the FFTs of the exponential and power-law fits to the experimental signal. We observe that not only the power-law model provides a better fit to the power spectrum of the data but also a sharply peaked spectrum, which is in principle key for improved precision and resolution. \\

\subsubsection{Qdyne with a single NV centre}

Qdyne experiments rely on individual storage of each single readout performed at a constant rate \cite{Schmitt2017, Degen2017}, in contrast to conventional measurements (as correlation spectroscopy) where the acquired data is averaged. In post-processing, the auto-correlation of the obtained time trace reveals the accumulated signal. For that reason a single dynamical decoupling measurement sequence is repeated continuously. The experiments were performed with NV centres at a moderate depth between 8 and 15\,nm. We performed six different Qdyne experiments spanning a total measurement time of $\approx 40$ days. The Fluka immersion oil is used for five and another perfluoropolyether oil (Fomblin Y, Sigma Aldrich 317993, viscosity of 1508\,cSt) for one measurement. XY8 dynamical decoupling is employed with different number of repetitions according to the depth of the NV centre where for all measurements the Larmor frequency is about $f_\mathrm{L} \approx 2\,\mathrm{MHz}$. Full details of the experimental parameters are given in subsection \ref{Appendix:Qdyne} in the Appendix. For each experiment, the resultant time-traces are divided into 15 minutes slices ($\sim 10^7$ measurements) for which the auto-correlation is calculated. Further noise reduction is achieved by averaging 20 of these auto-correlations. Frequency estimation is done by non-linear least squares fitting of the averaged auto-correlations to the modelling function Eq.~\eqref{eq:modelwithphase}, which compared to Eq.~\eqref{basicmodel}, contains a non-physical phase $\varphi$, included for reasons of stability in the numerical analysis. For each $C(t/T_D)$ model considered throughout, we perform a {\textit{local}} optimisation with fixed initial parameters, taking the initial frequency from FFT of the auto-correlation of the full time-trace, and a {\textit{global}} optimisation repeated 500 times with random initial parameters drawn from uniform distributions, in which the best fitting is determined by the highest $R^2$ (see \ref{Subsection:statistics} in Methods for full details). The latter analysis mimics the procedure in the case when FFT yields no results, e.g when the signal decay is too fast. 

\begin{figure}[t!]
\includegraphics[width=\columnwidth]{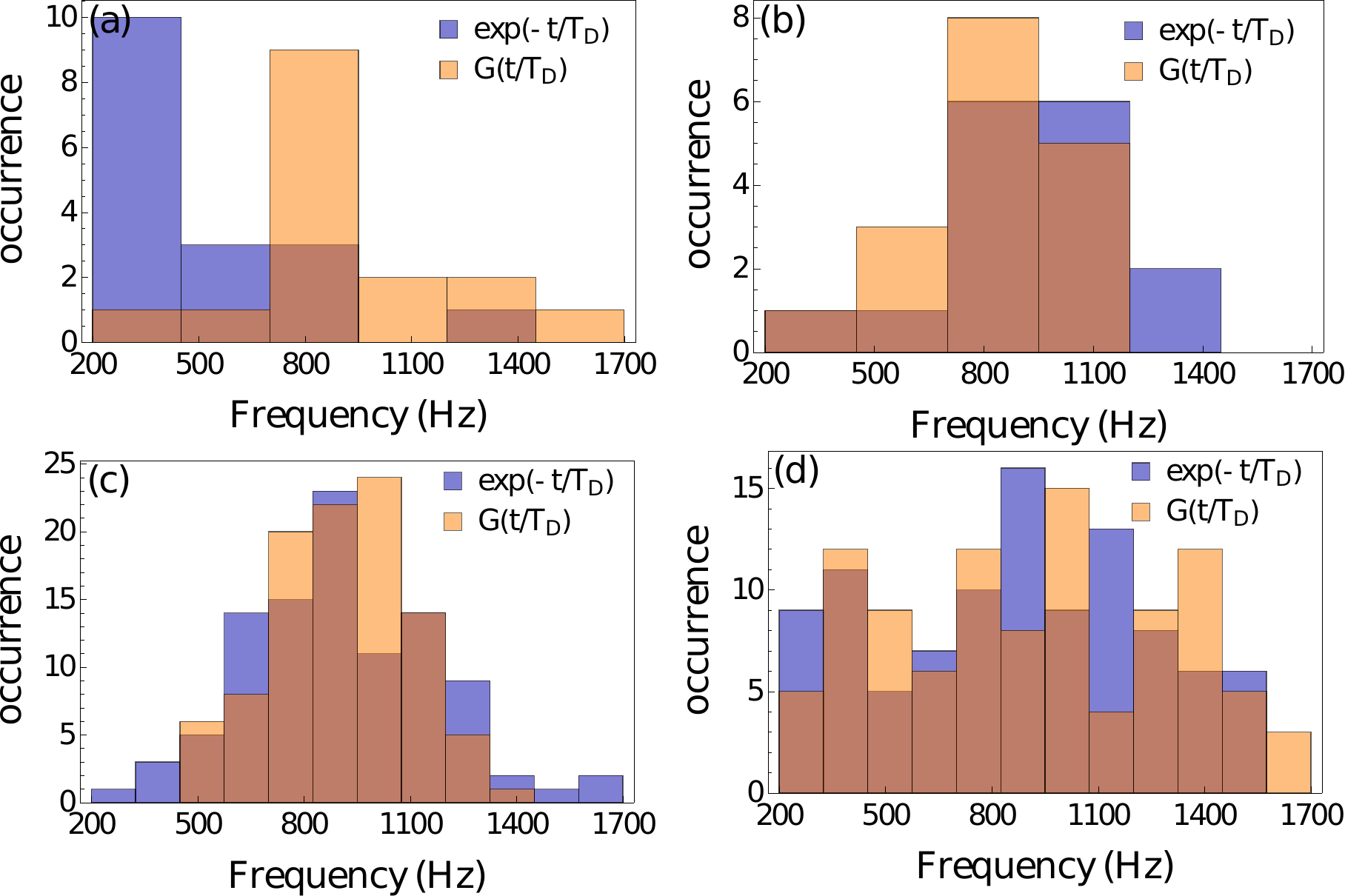}
    \caption{Histograms of frequency estimators. The data is obtained from non-linear least squares fitting of auto-correlations for experimental (upper row) and numerical (lower row) Qdyne time-traces. In blue, fitting function assumes exponential decay of correlations while in orange %the $C(t /T_D\gg 1) \sim (t/T_D)^{-3/2}$. 
    the power-law model is considered. Each estimator corresponds to the highest $R^2$ in 500 fittings of the same time-trace. In (a), the phase $\varphi$ is left as a free fitting parameter while in (b) it is kept fixed to $\varphi = 0$. (c) displays the results for numerical data generated for auto-correlations with power-law decay while in (d) the data is generated for exponentially decaying auto-correlations.
     }\label{Qdynefig1}
\end{figure}

Focusing on the first experiment (see Table~\ref{tab:Qdyne_details}), the upper row of Fig.~\ref{Qdynefig1} displays the histogram of 18 frequency estimators for global optimisation of each model, with $\varphi$ a free fitting parameter in Fig.~\ref{Qdynefig1}(a) and kept fixed ($\varphi = 0)$ in Fig.~\ref{Qdynefig1}(b). Results are consistent with the detection of a frequency $\delta \approx 900$ Hz, on a sample with estimated $T_D = 400\,\mathrm{\mu s}$, i.e. a frequency smaller than the noise bandwidth defined by the characteristic decay time of the auto-correlation. FFT analysis of the full signal's correlation (see Fig.~\ref{Fig:autocorr_calculation}) indicates a frequency centred at 970 Hz with a FWHM of 1205 Hz. In both instances of Fig.~\ref{Qdynefig1}, the root-mean-square error (rmse) of the estimator is smaller in the case of power-law correlations fitting as compared to exponential; 292 Hz vs 468 Hz in Fig.~\ref{Qdynefig1}(a) and 242 Hz vs 317 Hz in Fig.~\ref{Qdynefig1}(b), with the standard deviation showing similar scaling.

Further evidence of power-law correlations is provided by generating artificial time-traces numerically. We simulate Qdyne experiments with parameters similar to those in the experiment shown in the upper row of Fig.~\ref{Qdynefig1}, with two distinct noise models which produce two data sets with either exponential or power-law correlations as explained in \ref{Subsection:Numsim} in Methods. Each of the data sets is simultaneously analysed with both models, and the resulting histograms for the frequency estimators displayed in the lower row of Fig.~\ref{Qdynefig1}. In Fig.~\ref{Qdynefig1}(c), we show the results for data generated with power-law correlations, where a peak at a frequency $\approx$ 900 Hz can be estimated. Power-law model fitting yields a rmse of 195 Hz, while exponential fitting of the power-law auto-correlations is slightly wider at 270 Hz. Fig.~\ref{Qdynefig1}(d) displays the results corresponding to time-traces  generated with exponential correlations. Here, no frequency is detected and the rmse is 368 Hz for exponential fitting and 385 for the power-law model. Notice that the search region limit is 404 Hz for a flat histogram rmse, indicating that both estimators are essentially featureless. Together with the detection of a frequency below the noise bandwidth, our results are consistent with power-law correlations.

Note that in Fig.~\ref{Qdynefig1}(a), fitting to exponential correlations results in a histogram which goes to the extreme of the search region, signalling that the fitting generally fails. Only by reducing the number of parameters, restricting $\varphi$ to $0$, is the problem ameliorated. This does not happen for power-law model fitting. Results for local optimisation (displayed in Fig.~\ref{reslimitexperimentalappendix}) display similar behaviour and, for power-law correlations, there exists a trend in the rmse, which diminishes as the fitting model is refined, e.g. in local optimisation or with fixed parameters. Such trend is not found for exponential correlations fitting. That the fitting fails when an extra, meaningless parameter is added provides further evidence that it is the power-law model which best describes the experimental data.

\begin{figure}[t!]
    \includegraphics[width=\columnwidth]{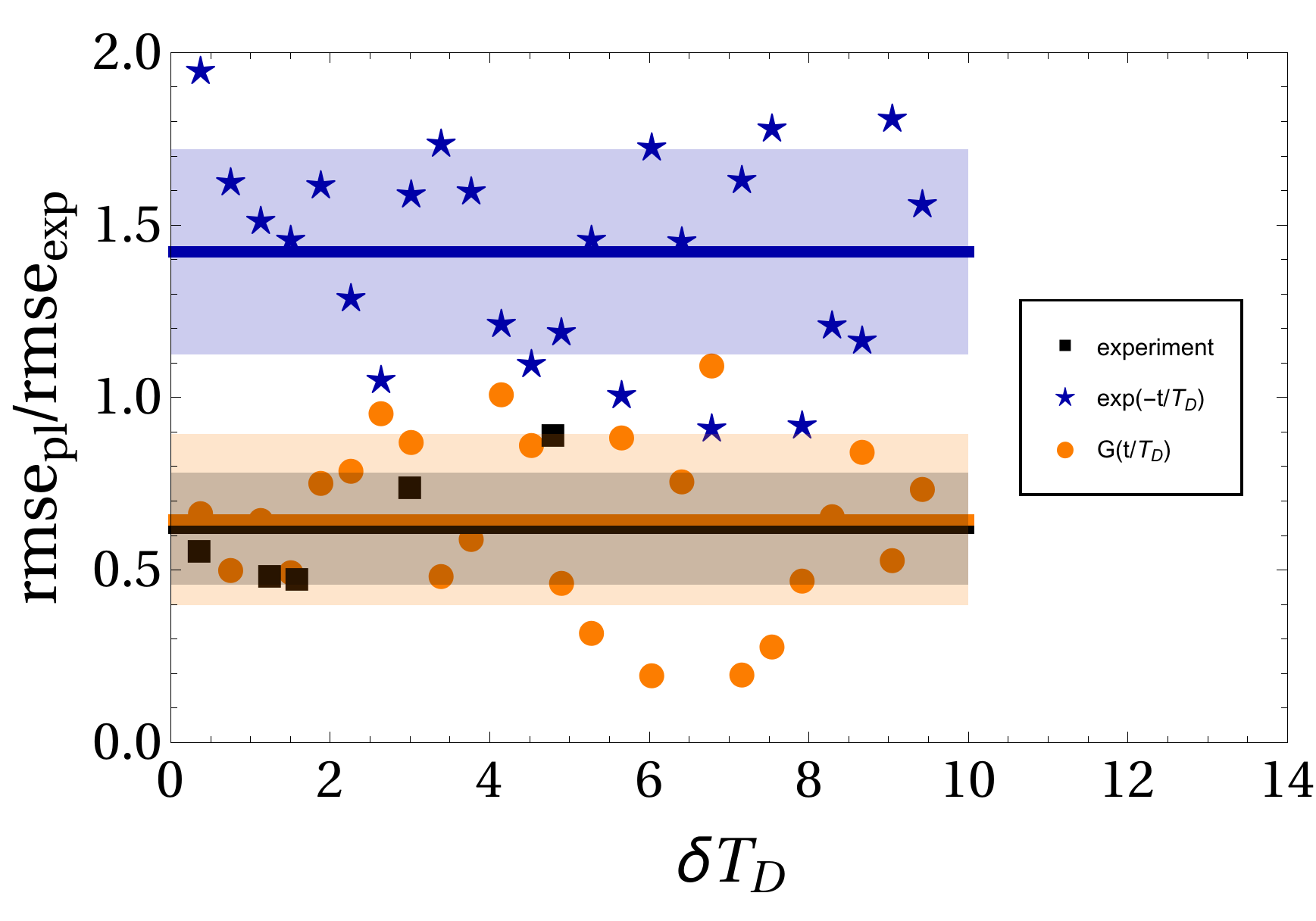}
    \caption{Root-mean-square error ratio between local optimisation for power-law and exponential model fittings. The ratio is plotted as a function of the frequency $\delta$ and the noise bandwidth defined by $T_D$. Each dot is calculated with the rmse of histograms as those shown in Fig.~\ref{Qdynefig1}. In black squares, experimental results. Orange circles display the ratio for data generated with power-law decay while blue stars correspond to data generated with exponential decay. Solid lines show the mean ratio for all dots while shaded areas correspond to the standard deviation for each mean. Note that experiment six in Table~\ref{tab:Qdyne_details} is not shown but it is considered for the mean ratio.}
     \label{Qdynefig2}
\end{figure}

We complete the analysis by considering the rmse ratio between power-law and exponential fittings. Here, most experiments have a frequency higher than the noise bandwidth (see Table~\ref{tab:Qdyne_details} for details) and therefore frequency estimation poses no problem with either model \cite{Oviedo2020}. Hence, we resort to compare which model provides the most accurate fitting according to a smaller rmse. Figure \ref{Qdynefig2} displays the rmse ratio for local optimisation of the experimental data, with the dashed line showing the average ratio standing at 0.62 $\pm$ 0.16, which means that power-law fitting of the experimental data has an average improvement of 40 $\%$ of the rmse with respect to exponential fitting. For global optimisation the improvement is 20 $\%$ (see Fig.~\ref{Qdynefigratioglobal}). A numerical analysis of artificial time-traces generated with either correlation model and analysed with both models shows the striking difference existing when analysing the data with the incorrect model. For data generated with power-law correlations the root-mean-square error (rmse) ratio between local
optimisation for power-law and exponential model fittings is 0.65 $\pm$ 0.27 while for data generated with exponential correlations the ratio is 1.42 $\pm$ 0.29, further confirming the experimental results of detection of power-law decay of correlations.

%%%%%%%%%%%%%%%%%%%%%%%%%%%%%%%%%%%%%%%%%%%%%%%%%%%%%%%%%%%%%%%%%%%%%%%%%%%%%%%%%%%%%%%%%%%%%%%%%%%%%%%%%%%%%%%%%%%%%%%%%%%%%%%
%%%%%%%%%%%%%%%%%%%%%%%%%%%%%%%%%%%%%%%%%%%%%%%%%%%%%%%%%%%%%%%%%%%%%%%%%%%%%%%%%%%%%%%%%%%%%%%%%%%%%%%%%%%%%%%%%%%%%%%%%%%%%%%
\subsubsection{Diffusion coefficient estimation}

Estimating the diffusion coefficient on the micro- and nano-scales is an important but challenging task, and current techniques are prone to errors \cite{Farida2021}. Using NV centres to calculate $D$ from the characteristic time $T_D$ in an exponential envelope of correlations leads to inaccuracies, as this model disregards correlations beyond the distance $d$, which nonetheless carry diffusion information, as shown by a more rigorous modelling \cite{Cohen2020}. The ability to measure power-law correlations, as we have done throughout this article, opens up the possibility to estimate $D$ more accurately.  

The diffusion time $T_D$ is defined as the characteristic time to diffuse at a distance $d=\sqrt{D T_D}$. For a hemispherical sensing volume, the number of nuclei is $N \propto d^3$, then $B_\mathrm{rms} \propto \frac{\sqrt{N}}{d^3} \propto d^{-3/2}$ and we have that $d \sim \brms^{-2/3}$. Thus, the procedure to obtain $D$ shall be as follows. The $\brms$ can be independently estimated from, e.g., power spectrum measurements (as we do here) or rapid Ramsey measurements; hence the depth $d$ of the NV centre can be obtained. $T_D$ is then estimated from a fitting procedure as described in the previous sections, with the $\brms$ a fixed parameter. From these, the diffusion coefficient $D$ can be obtained. 

Following this procedure, we estimate $D$ for the correlation spectroscopy measurements with single NV centres shown in Fig.~\ref{Fig:CS_NV4_v3}. For the exponential model we get $D_\mathrm{exp}\approx 6.65~\times 10^{-13}\,\mathrm{m^2\,s^{-1}}$, with a 95\% confidence interval $(5.57 - 8.25 )\times 10^{-13}~\mathrm{m^2\,s^{-1}}$,
while the power-law model results in $D_\mathrm{pl} \approx 4.33~\times 10^{-13}\,\mathrm{m^2\,s^{-1}}$ with a 95\% confidence interval $(2.95 - 8.15)\times 10^{-13}~\mathrm{m^2\,s^{-1}}$. The estimate of the diffusion coefficient of the oil in the experiment, based on its viscosity characteristics is of the order of magnitude \cite{Pham2016} $D \approx 6~\times 10^{-13}\,\mathrm{m^2\,s^{-1}}$, which is within the confidence intervals of the estimates of both models. 
%The discrepancy arising from the differences between the two correlation's models, e.g., in the fitting envelopes displayed in Figs.~\ref{Fig:CS_NV4_v3} or \ref{Fig:CS_NVision}. 
While the reported confidence intervals are rather large, they could be substantially decreased in future experiments, specifically designed to improve the efficiency of the procedure, e.g., by improving the signal-to-noise ratio and obtaining a clear signal at even longer times, %e.g., $\approx 10 T_{D}$, 
as expected theoretically from the power-law model. 
Investigation of surface effects on diffusion are also envisaged, e.g., by probing it with NV centres at different depths with surface effects expected to be more pronounced with shallower NV centres.

%%%%%%%%%%%%%%%%%%%%%%%%%%%%%%%%%%%%%%%%%%%%%%%%%%%%%%%%%%%%%%%%%%%%%%%%%%%%%%%%%%%%%%%%%%%%%%%%%%%%%%%%%%%%%%%%%%%%%%%%%%%%%%%
%%%%%%%%%%%%%%%%%%%%%%%%%%%%%%%%%%%%%%%%%%%%%%%%%%%%%%%%%%%%%%%%%%%%%%%%%%%%%%%%%%%%%%%%%%%%%%%%%%%%%%%%%%%%%%%%%%%%%%%%%%%%%%%

%%%%%%%%%%%%%%%%%%%%%%%%%%%%%%%%%%%%%%%%%%%%%%%%%%%%%%%%%%%%%%%%%%%%%%%%%%%%%%%%%%%%%%%%%%%%%%%%%%%%%%%%%%%%%%%%%%%%%%%%%%%%%%%
%%%%%%%%%%%%%%%%%%%%%%%%%%%%%%%%%%%%%%%%%%%%%%%%%%%%%%%%%%%%%%%%%%%%%%%%%%%%%%%%%%%%%%%%%%%%%%%%%%%%%%%%%%%%%%%%%%%%%%%%%%%%%%%
\section{Discussion}\label{Section:Discussion}
%%%%%%%%%%%%%%%%%%%%%%%%%%%%%%%%%%%%%%%%%%%%%%%%%%%%%%%%%%%%%%%%%%%%%%%%%%%%%%%%%%%%%%%%%%%%%%%%%%%%%%%%%%%%%%%%%%%%%%%%%%%%%%%
%%%%%%%%%%%%%%%%%%%%%%%%%%%%%%%%%%%%%%%%%%%%%%%%%%%%%%%%%%%%%%%%%%%%%%%%%%%%%%%%%%%%%%%%%%%%%%%%%%%%%%%%%%%%%%%%%%%%%%%%%%%%%%%

An exponential modelling for correlation's decay is the natural assumption for nano-NMR experiments. Yet this hypothesis has profound implications on the possible applications of any such experiments. In particular, exponential correlations lead to Lorentzian lineshapes for which spectral resolution is fundamentally limited. 
It is possible, however, to refine the original assumption by carefully examining the microscopic diffusion process which is the cause of correlations decay. Diffusion causes power-law decay of correlations at long times for the dipole-dipole interaction that is characteristic for the NV centre nano-NMR spectrometer \cite{Cohen2020}. This has substantial consequences, as spectral lineshapes become sharp-peaked, for which in theory there is no limit to resolution \cite{Oviedo2020}. Yet, measuring deviations from the exponential paradigm and utilising them for improved spectral analysis is a challenging task due to experimental noise. Furthermore, by considering the full interaction model for correlations rather than the truncated exponential model, it shall be possible to measure more precisely the diffusion coefficient $D$ in microfluids, a long sought goal, where current methods have only limited accuracy \cite{Eggeling2009,Heinemann2012}.

By probing the auto-correlation of the dipole-dipole interaction between NV centres and nuclear spins, and prolonging the measurement time to several times the characteristic exponential decay time, we show strong evidence supporting a power-law decay of correlations as they provide a better fit to the experimental results. % than the exponential model. 
Furthermore, with the Qdyne data frequency estimation yields a results, which is 40$\%$ more precise than with the exponential model.
We can also exclude surface effects causing a long correlation tail by detecting the signal created by fluorine nuclei lying on top of a proton layer adhered to the diamond surface, in which such effects originate \cite{Wrachtrup2015}. Additionally, we show that Qdyne measurements auto-correlations are compatible with the same power-law decay to the extent that in the regime where exponential correlations limit frequency resolution, a frequency can still be detected. Finally, we rule out the possibility of exponential correlations by showing with numerical analysis that such correlations in the same regime would lead to featureless histograms or altogether wide spectral lines in which frequency estimation is not possible. 

Our results offer a way to significantly improve spectral resolution, and enable broad applications of nano-NMR with NV centres. The power-law scaling of correlations which originates in the dipole-dipole interaction between sensor and sample is key to this result, extending the applicability of our results to any quantum sensor based on this interaction, such as Rydberg atoms, squid based sensors, molecular quantum sensors, or alternative colour centres such as silicon carbide \cite{Yu2021}. In addition, the ability to measure accurately correlations of nano-sized fluid samples, opens the door to study flow properties at these scales. It has been recently shown that single trajectory auto-correlations differ from ensemble averaged (multiple trajectories) auto-correlations for anomalously diffusing fluids \cite{Metzler2019}, which also show power-law scaling characterised by critical exponents \cite{Sadegh2014,Leibovich2016}. These single auto-correlations contain crucial information about the sample fluid which might be missed on typical ensemble averages \cite{Krapf2019}. Our results show that it is possible to directly access the microscopic behaviour that underpins the properties of nano-fluids, with applications in a wide variety of fields.

%%%%%%%%%%%%%%%%%%%%%%%%%%%%%%%%%%%%%%%%%%%%%%%%%%%%%%%%%%%%%%%%%%%%%%%%%%%%%%%%%%%%%%%%%%%%%%%%%%%%%%%%%%%%%%%%%%%%%%%%%%%%%%%
%%%%%%%%%%%%%%%%%%%%%%%%%%%%%%%%%%%%%%%%%%%%%%%%%%%%%%%%%%%%%%%%%%%%%%%%%%%%%%%%%%%%%%%%%%%%%%%%%%%%%%%%%%%%%%%%%%%%%%%%%%%%%%%
\section{Methods}\label{Section:Methods}
\subsection{Samples}\label{Subsection:samp}
All experiments are carried out with an isotopically enriched (99.999\,$\%$ $^{12}$C) diamond sample grown by chemical vapour deposition. The diamond  substrate %, purchased from Element Six Inc., 
with natural abundance of $^{13}$C was equipped with a $99.999\,\%$ $^{12}$C enriched homoepitaxially grown diamond film with a thickness of about 150\,nm in a home-built plasma enhanced chemical vapour deposition growth reactor \cite{osterkamp2019engineering,silva2010microwave,findler2020indirect} using 99.999\,$\%$ enriched \textsuperscript{12}CH\textsubscript{4} gas (Cambridge Isotope Laboratories) at a concentration of 0.2\,$\%$ with respect to hydrogen. 

Isolated shallow NV centres were then created by ion implantation with $^{15}$N$^+$ at a dose of $5 \times 10^{8}\,\mathrm{N^+\,cm^{-2}}$, using an acceleration energy of 2\,keV (correlation spectroscopy), and 2.5\,keV (Qdyne). Additionally, for Qdyne measurements utilising deeper NV centres, an ion dose of $1 \times 10^{11}\,\mathrm{N^+\,cm^{-2}}$ at an acceleration voltage of 2.5\,keV was used, followed by processing the diamond with the indirect overgrowth method according to Ref. \cite{findler2020indirect}, burying the NV centres deeper and thus shielding them from noise sources at the surface of the diamond. To heal resulting radiation damage, mobilise vacancies, and eventually create the desired NV centres, the diamond samples are annealed in a home-built UHV furnace at $1000\,^\circ\mathrm{C}$ for 3 hours, while ensuring extremely low process pressures $< 1 \times 10^{-7}$\,mbar \cite{lang2020long}.

The NV centre ensembles were created by implanting $^{15}$N$^+$ ions with a dose of \SI{1e12}{\text{N\textsuperscript{+}}/\centi \metre \squared} and an energy of \SI{2.5}{\kilo \electronvolt}, followed by annealing as described for the single NV centres. The NV ensemble creation yield was determined to be \SI{1 \pm 0.1}{\percent} which corresponds to an average NV concentration of \SI{60}{ppb} taking into account the depth distribution of implanted nitrogen in reference \cite{findler2020indirect}. 
After annealing, the diamond is boiled in a 1:1:1 mixture of sulphuric (97$\%$), perchloric (70$ \%$) and nitric acid (65$\%$) at $200 \,^\circ\mathrm{C}$ in a microwave reactor system (MWT AG, type ETHOS Lab) for 30 minutes to remove any (graphitic) residues from the surface.

\subsection{Power spectrum measurement}\label{Subsection:PS}

We perform power spectrum measurements to determine the $\brms$ of the nuclear spins at the NV centre position \cite{Pham2016}. This is used to determine the initial contrast of the auto-correlation signal and fix the value as in Figure \ref{Fig:CS_NV4_v3}(b) and \ref{Fig:CS_NVision}(b). We consider a two-state system, which is prepared initially in state $\ket{0}$, e.g. by optical pumping. The power spectrum measurement consists of the microwave pulse sequence $\pi/2(x)$ pulse -- dynamical decoupling -- $\pi/2(\mp x)$ pulse, where the coordinate in parentheses indicates the axis of rotation. For two alternating measurements, which differ by the axis of rotation of the last $\pi/2$-pulse, we take their signal difference. The advantage of using the alternating sequence is reduction of the effect of unwanted laser power fluctuations. 

During the DD sequence a phase $\Phi$ is accumulated due to the Larmor precession of the nuclear spins. $\Phi$ follows a Gaussian distribution where the expectation of $\Phi$ is zero. Its variance depends on the strength of the interaction between the NV centre and the nuclear spins in the sensing volume, the filter function of the applied DD sequence and the interaction time. The expected value of the phase variance in case of statistical polarisation is given by $\langle\Phi^2\rangle=\Phi_{\text{rms}}^2$ (see Appendix \ref{Section:SP_modeling}) 
\begin{align}\label{Eq:phi_rms}
\Phi_{\text{rms}}=\frac{2}{\pi}\gamma_e B_{\text{rms}} N\tau
~\text{sinc}{\left(N(\pi-\omega_{\text{L}}\tau)\right)},
\end{align}
where $\gamma_e$ is the gyromagnetic ratio for the NV electron spin, $B_{\text{rms}}$ is the root-mean-square of the magnetic field of the statistically polarised sample, $N$ is the number of the DD pulses, $\tau$ is their pulse separation and $\omega_{\text{L}}$ is the Larmor frequency of the sensed spins (in angular frequency units). This allows us to estimate $B_{\text{rms}}$ and the corresponding depth of the NV centre \cite{Pham2016}.

\subsection{Correlation spectroscopy}\label{Subsection:CS}

Next, we analyse the expected measurement results with correlation spectroscopy. Again the two-state system is considered, which is prepared initially in state $|0\rangle$. We perform the correlation spectroscopy measurement, which consists of two DD blocks, separated by a waiting time $T$. During the waiting time, the information about the accumulated phase is mapped onto a population difference, so it cannot exceed $T_1$ of the system. The pulse sequence is
\begin{gather}
\pi/2(x)- \text{DD sequence} -  \pi/2(y) \notag\\
- \text{ waiting time $T$ } -  \notag\\
\pi/2(x)- \text{DD sequence} -  \pi/2(\pm y),
\end{gather}
where the two versions of the last pulse $\pi/2(\pm y)$ give the alternating projections on the $|0\rangle$ and $|1\rangle$ states, respectively. The signal is the  difference between these two and is given by
\begin{align}
c_{\text{cs}}&=
c_\text{max}\sin{\left(\Phi_1\right)}\sin{\left(\Phi_2\right)},
\end{align}
where $\Phi_1$ and $\Phi_2$ are the accumulated phases during the first and the second DD sequences, $c_\mathrm{max}$ is the maximum readout contrast (see Appendix~\ref{Section:Readout} for details), and we can neglect the effect of decoherence and relaxation because the duration of each dynamical decoupling sequence $\tau N \ll T_2$ and the waiting time $T \ll T_1$. 

Averaging multiple measurement readouts leads to
\begin{align}
\overline{c}_{\text{cs}}=
c_\text{max}\left\langle\sin{\left(\Phi_1\right)}\sin{\left(\Phi_2\right)}\right\rangle
\approx c_\text{max}\left\langle\Phi_1\Phi_2\right\rangle,
\label{eq:corrspec_signal}
\end{align}
where the last approximation is valid only for small $\Phi_k,~k=1,2$. The phases $\Phi_k$ follow the same Gaussian distribution as in the power spectrum measurement, i.e. centred at zero with a variance $\Phi_{\text{rms}}^2$, and we obtain (see Appendix~\ref{Section:SP_modeling})
\begin{align}
\left\langle\Phi_1\Phi_2\right\rangle=\Phi_{\text{rms}}^2 \cos{(\omega_{\text{L}} t)} C(t),
\label{signal_shape}
\end{align}
where $t = N\tau+T$ is the time separation between the beginnings of the two DD sequences and $\omega_{\text{L}}=2\pi f_{\text{L}}$ with $f_{\text{L}}$ again the Larmor frequency of the sensed spins (or the respective undersampling frequency). In our analysis the correlation $\left\langle B(t^\prime)B(t^\prime+t)\right\rangle\approx B_{\text{rms}}^2 \cos{(\omega_{\text{L}}t)}C(t)$, where the correlation function $C(t)$ can be exponential \cite{Pham2016} or have a power-law decay for long times \cite{Cohen2020,Oviedo2020}.

\subsection{Quantum heterodyne detection}\label{Subsection:Qdyne}

Lastly, we focus on the Qdyne protocol \cite{Schmitt2017}. As before, the initial state of the NV centre is the $\ket{0}$ state. In this case, the basic building block of the protocol is a DD sequence encased in between two $\pi/2$ pulses as 
\begin{gather}
\pi/2(x)- \text{DD sequence} -  \pi/2(y).
\end{gather}
The salient feature of the Qdyne protocol is that the projection of the NV state through the second $\pi/2$ pulse, is followed not only by state readout/re-initialisation but also by a fixed delay time that defines the sampling frequency with which measurements are repeated, thereby recording information about the phase of the target signal, a crucial point that permits correlating all measurement outcomes in post-processing. 

No alternating measurement protocol is used here. The signal for a given measurement at time $t$ in a Qdyne measurement sequence is
\begin{equation}
c_\mathrm{Qd} = \frac{c_{\text{max}}}{2} \sin(\Phi_t). 
\label{eq:qdyne_signal}
\end{equation}
In post-processing of the data the auto-correlation of the recorded time trace is calculated. The correlation between any two readout pairs that are separated by $t_n = n T_\mathrm{Qd}$, where $T_\mathrm{Qd}$ is the sequence length of a single measurement, is then
\begin{equation}
\begin{aligned}
\bar{c}_{Qd} &= \avg{c_{\mathrm{Qd},t}\, c_{\mathrm{Qd}, t+t_n}} = c_{\text{max}}^2\langle\sin(\Phi_t)\sin(\Phi_{t+t_n})\rangle \\
&\approx c_{\text{max}}^2\langle\Phi_t\Phi_{t+t_n}\rangle.
\end{aligned}
\end{equation}
With $\Phi_t = \Phi_1$ and $\Phi_{t+t_n} = \Phi_2$ the same signal as in correlation spectroscopy \eqref{eq:corrspec_signal} is measured.

\subsection{Statistical analysis}\label{Subsection:statistics}

Parameter estimation is done by numerically fitting the experimental auto-correlation functions to the theoretical mode
\begin{equation}
    a_0+a_1\cos{(\delta t+\varphi)}C(t/T_{D}),\label{eq:modelwithphase}
\end{equation}
where $a_0$ is a signal offset, $a_1\sim \Phi_{\text{rms}}^2$ is the signal amplitude, $\delta$ is the undersampling frequency, $\varphi$ is the signal phase, and $T_{D}$ is the diffusion time. In the fits of Figs.\,\ref{Fig:CS_NV4_v3} and\;\ref{Fig:CS_NVision} the parameter $a_1\sim \Phi_\mathrm{rms}^2$ is fixed and estimated from an independent power spectrum measurement.
Note that we have included an artificial phase $\varphi$. This is necessary for correlation spectroscopy experiments as the undersampling frequency $\delta$ is typically much smaller than the actual Larmor frequency $\omega_{\text{L}}$, which can result in a non-zero phase, e.g., when the first sampling time is smaller than the oscillation period at $\delta$. The reason is purely numerical for Qdyne as the fitting algorithm is prone to crashing, especially when considering the power-law correlations with Eq.~\eqref{Cohenian}. Then, the phase aids in adding stability and thus saving computational time. In all of the instances where we include it as a free parameter, we check that the estimation result for $\varphi$ is either $0$ or $2\pi$ within numerical precision. 
For the correlation function decay described by the envelope $C(t/T_D)$ we consider either that (exponential model) 
\begin{equation}
    C(t/T_D) = \exp(-t/T_D), \label{expmodel}
\end{equation}
or (power-law model) as in \cite{Cohen2020}
\begin{equation}
\begin{aligned}
G(z) = \frac{4}{\sqrt{\pi}} \Bigg(
& z^{-\frac{3}{2}} - \frac{3}{2}z^{-\frac{1}{2}} + \frac{\sqrt{\pi}}{4} + 3\sqrt{z} - \frac{3\sqrt\pi}{2}z + \\
& \sqrt{\frac{\pi}{z}}{\text{erfc}}\Big(z^{-\frac{1}{2}}\Big)\exp{z^{-1}} \times \\
& \Big( - z^{-\frac{3}{2}} + z^{-\frac{1}{2}} - \frac{7}{4}\sqrt{z} + \frac{3}{2}z^{+\frac{3}{2}} \Big)  \Bigg),
\label{Cohenian}
\end{aligned}
\end{equation}
with $z = t/T_D$ and $C(z) = G(z)$ to distinguish it from other possible power-law models. The latter is shown to fall off as $C(z\ll1) \propto z^{-3/2}$, for which reason we call it the power-law model. 

The fitting procedure utilises a non-linear least squares algorithm with finite-difference estimation of the gradient. In correlation spectroscopy data, due to measurement vectors being indivisible, we compare goodness of fitting for each of the two correlation models, using the $R^2$ as a figure of merit. For each signal and model, the fitting procedure is repeated 500 times and the best fitting according to the highest $R^2$ is selected. The procedure is akin to the global optimisation described below for the Qdyne data.

Qdyne data analysis deserves a more in-depth explanation. Qdyne experiments are performed sequentially, therefore, each experiment yields a single time-trace of measurements spanning the total duration of the experiment. Here, we describe the processing of the data corresponding to the experiment shown in Fig.~\ref{Qdynefig1}(a) and (b), as the rest are analogous. In this case, the total measurement duration is 90 hours, while each single measurement requires 49.740 $\mu$s. Contrary to the case of correlation spectroscopy, where the data recording procedure did not allow to slice the time-traces, in Qdyne we can partition the data in arbitrarily short vectors. For reasons of setup stability the slices are of 15 minutes worth of measurements. In between two of these 15-minute time-traces the NV centre is optically refocused. However, the resulting auto-correlations are too noisy for adequate fitting. Thus, we compromise between eliminating noise by averaging several auto-correlations and having sufficient statistics to build a meaningful histogram, i.e. to perform Bayesian analysis. For Fig.~\ref{Qdynefig1}(a) and (b) that means averaging 20 auto-correlations from 15-minute time-traces, resulting in a total of 18 auto-correlations with which to perform statistical analysis. Each of the 18 resulting averaged auto-correlations is fitted to the model as explained above, yielding a total of 18 frequency estimators that compose the shown histograms. 

In the case of local optimisation each of the 18 auto-correlations is fitted once to the model Eq.~\eqref{basicmodel}, with input parameters taken from the FFT of the total auto-correlation calculated over the full 90 hours time-trace for the frequency, and amplitude and decay time taken directly from the full auto-correlation. For global optimisation, the fitting procedure is repeated 500 times, each time with different initial parameters randomly drawn from uniform distributions of size equal to the allowed searching regions for the fitting procedure (e.g. from 200 to 1800 Hz for the frequency estimator). This procedure is repeated for each of the 18 time-traces and for each case the frequency of the highest $R^2$ fitting results in the estimator considered for statistics. In all cases the figure of merit is the root-mean-square error of the resulting histogram of estimators.

\subsection{Numerical simulations}\label{Subsection:Numsim}

Qdyne signals are generated numerically employing data from molecular dynamics simulations \cite{Oviedo2020,Cohen2020b}.
%two alternative, complementary methods \cite{Oviedo2020}. 
We utilise molecular dynamics simulations data of $N \approx 46\times10^3$ dipolar particles within a simulation box of size $L_{x,y,z} = (50,50,24)\,\mathrm{nm}$ with an NV centre located at depths ranging from 0.3 to 12\,nm. The particles within the box diffuse according to a Lennard-Jones fluid with normalised parameters $\epsilon = \sigma = 1$ (where $\epsilon$ is the depth of the potential while $\sigma$ is the distance at which the potential is zero), starting from an initial thermal state at temperature $T=\frac{\kappa_B T(K)}{\epsilon}$ \cite{Cohen2020}. The magnetic signal is calculated as the magnetic field induced by the particles at the NV position. The data thus generated proves to have no trends and the standard deviation is scale invariant, i.e. depth independent, meaning that noise data generated for one depth can be scaled through the corresponding $T_D$ to suit a different depth \cite{Oviedo2020}. Therefore, we are able to use all depths, thus, increasing the statistics. %This data is then re-scaled appropriately to reflect a single relevant depth. 
The resultant time-traces contain the power-law noise model with correlations scaling as $C(t/T_D \gg 1) \sim (t/T_D)^{-3/2}$ and thus we use them directly to generate Qdyne time traces associated with this model. Exponential correlations are simulated by fitting the molecular dynamics data to Orstein-Uhlenbeck noise which process exponential decay, and using such fitting as the noise source.

%\section{Acknowledgments}
%S.O.C. acknowledges the support from the \textsl{Fundaci\'on Ram\'on Areces} postdoctoral fellowship (XXXI edition of grants for Postgraduate Studies in Life and Matter Sciences in Foreign Universities and Research Centres). A.V.S. acknowledges the support from the \textsl{European Union’s Horizon 2020 research and innovation program} under the Marie Skłodowska-Curie Grant Agreement No. 766402. N.Sta. acknowledges support from the \textsl{Bosch-Forschungsstiftung}. D.C. acknowledges the support of the \textsl{Clore Scholars Programme} and the \textsl{Clore Israel Foundation}. 
%This work was supported by the European Union’s Horizon 2020 research and innovation program under grant agreement No 820394 (ASTERIQS), DFG (CRC 1279 and Excellence cluster POLiS), ERC Synergy grant HyperQ (Grant No. 856432), BMBF and VW Stiftung. A.R. acknowledges the support of ERC grant QRES, project number 770929, grant agreement number 667192 (Hyperdiamond), ISF and the Schwartzmann university chair.


\section*{Data Availability}
The data that support the findings of this study are available from the corresponding author upon reasonable request. 

\section*{Code Availability}
The code used for obtaining the presented numerical results is available from the corresponding author upon reasonable request. 

\section*{Acknowledgments}
S.O.C. acknowledges the support from the \textsl{Fundaci\'on Ram\'on Areces} postdoctoral fellowship (XXXI edition of grants for Postgraduate Studies in Life and Matter Sciences in Foreign Universities and Research Centres). A.V.S. acknowledges the support from the \textsl{European Union’s Horizon 2020 research and innovation program} under the Marie Skłodowska-Curie Grant Agreement No. 766402. N.Sta. acknowledges support from the \textsl{Bosch-Forschungsstiftung}. D.C. acknowledges the support of the \textsl{Clore Scholars Programme} and the \textsl{Clore Israel Foundation}. 
This work was supported by the European Union’s Horizon 2020 research and innovation program under grant agreement No 820394 (ASTERIQS), DFG (CRC 1279 and Excellence cluster POLiS), ERC Synergy grant HyperQ (Grant No. 856432), BMBF and VW Stiftung. A.R. acknowledges the support of ERC grant QRES, project number 770929, grant agreement number 667192 (Hyperdiamond), ISF and the Schwartzmann university chair.

\section*{Author contributions}
A.R., D.C. and F.J. conceived the idea. N.Sta. did the Qdyne measurements with single NV centres. A.V.S and G.G. performed the correlation spectroscopy measurements with single NV centres, while D.D., T.U., N.Str. and P.N. did the corresponding experiments with ensembles of NV centres. A.M. and J.S. helped with the integration of hardware and software. A.V.S., G.G., N.Sta., and S.O.C. did the analysis of the experimental results and S.O.C. did the simulation for the numerical data. C.F. and J.L. provided the diamond samples for the measurements. A.V.S., G.G., N.Sta., and S.O.C. prepared the manuscript. All authors read and contributed to the paper. A.R., I.S., P.N. and F.J. supervised the project. N.Sta., A.V.S., S.O.C. and G.G. contributed equally.

\clearpage
\newpage

\appendix

%%%%%%%%%%%%%%%%%%%%%%%%%%%%%%%%%%%%%%%%%%%%%%%%%%%%%%%%%%%%%%%%%%%%%%%%%%%%%%%%%%%%%%%%%%%%%%%%%%%%%%%%%%%%%%%%%%%%%%%%%%%%%%%
%%%%%%%%%%%%%%%%%%%%%%%%%%%%%%%%%%%%%%%%%%%%%%%%%%%%%%%%%%%%%%%%%%%%%%%%%%%%%%%%%%%%%%%%%%%%%%%%%%%%%%%%%%%%%%%%%%%%%%%%%%%%%%%

\section{Quantum state readout}\label{Section:Readout}

\subsection{Power spectrum analysis}\label{Subsection:PSappendix}

We consider a two-state system, which is prepared initially in state $\ket{0}$, e.g., by optical pumping.
In all measurement schemes, we find the population of the NV centre via fluorescence collection during laser excitation, which relates the average number of detected photons to the population of the system because the fluorescence rate is higher for the NV centre being in the spin state $\ket{0}$. 
To reduce laser power fluctuations, the photoluminescence (PL) of the NV centres at the first few hundred nanoseconds (signal window) of the laser pulse are normalised with the reference PL at the end of the optical pumping period (reference window; usually last 2\,$\mu$s of a 5\,$\mu$s long laser pulse). The signal window is adjusted such that highest SNR can be achieved. The fluorescence signal for readout of the state $\ket{i}$ ($i = 0,1$) is $F_i = \frac{\eta_i}{\eta_\mathrm{ref}}$ where $\eta_i$ is the photon count rate during the signal window if the NV centre is in state $\ket{i}$ and $\eta_\mathrm{ref}$ is the photon count rate during the reference window that is independent of the spin state after long enough optical pumping. The maximum contrast is then $c_\mathrm{max} = F_0 - F_1 = \frac{\eta_0 - \eta_1}{\eta_\mathrm{ref}}$ and the readout signal for an arbitrary spin state with populations $p_0$ and $p_1$ is given by $F = p_0 F_0 + p_1 F_1 = \frac{p_0 \eta_0 + p_1 \eta_1}{\eta_\mathrm{ref}}$. 

In the experiments usually a maximum contrast $c_\mathrm{max} \approx 30\,\%$ is obtained. %However, in the experiment with the single NV centre the contrast has dropped over the measurement time due to charge state instabilities most probably favored by long laser illumination. 

We describe now the power spectrum measurement, which consists of the microwave pulse sequence $\pi/2(x)$ pulse - dynamical decoupling - $\pi/2(-x)$ pulse. This sequence keeps the system in state $\ket{0}$ in case of no phase accumulation during dynamical decoupling. 
The density matrix after the sequence is given by
\begin{align}
\rho_{\text{ps}}
&=\left[ \begin{array}{cc} \cos\left(\frac{\Phi}{2}\right)^2 & \frac{\sin\left(\Phi\right)}{2} \\
\frac{\sin\left(\Phi\right)}{2}  & \sin\left(\frac{\Phi}{2}\right)^2 \end{array}\right],
\end{align}
where $\Phi$ is the accumulated phase during the DD sequence and the population of state $\ket{0}$ and $\ket{1}$ is given by $p_0 =  \cos\left(\frac{\Phi}{2}\right)^2$ and $p_1 =  \sin\left(\frac{\Phi}{2}\right)^2$, respectively. 

\begin{figure}[t!]
\includegraphics[width=\columnwidth]{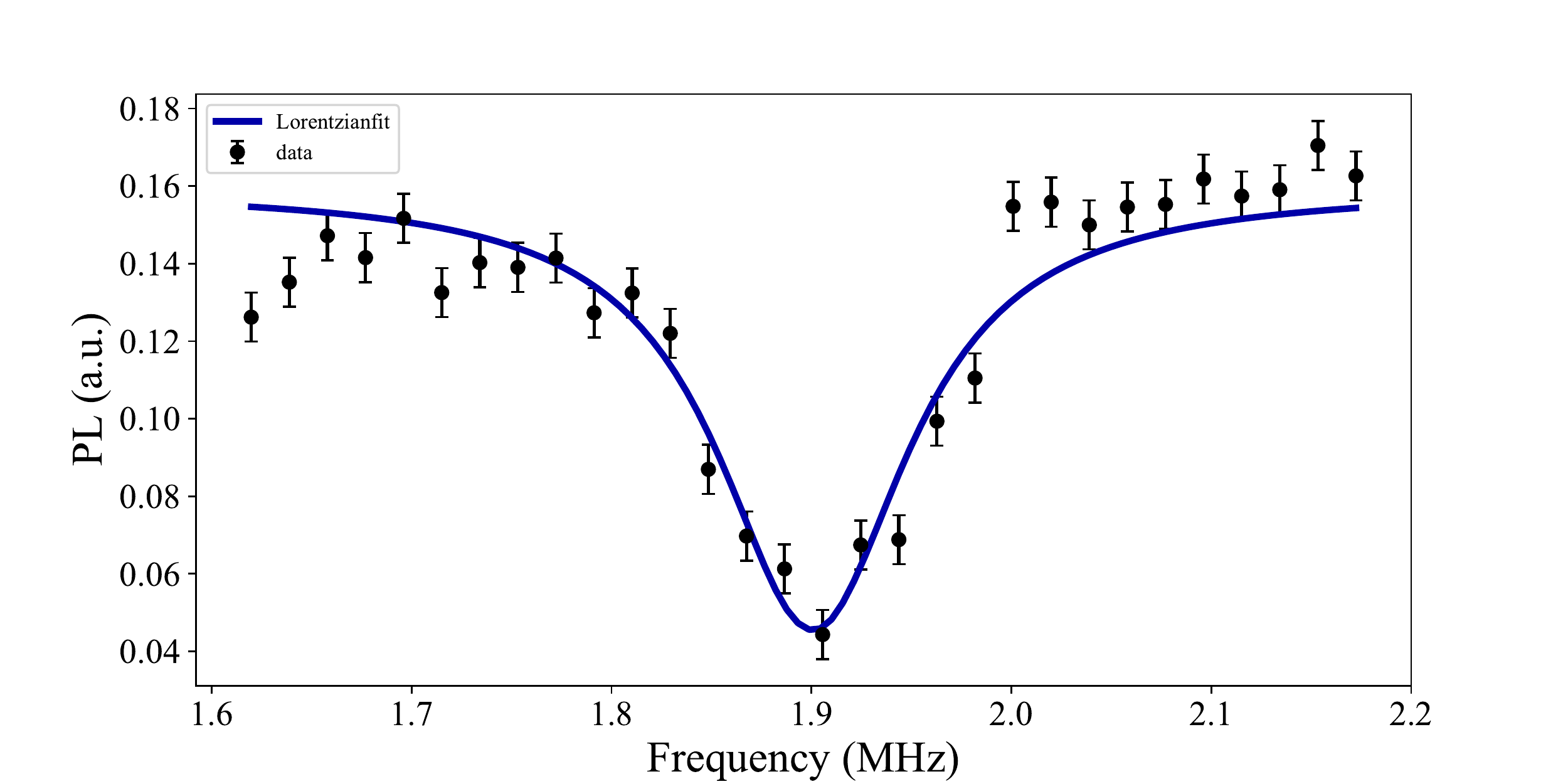}
\caption{Hydrogen power spectrum obtained with KDD4-2. The experimental data points are fitted with a Lorentzian fit (blue line). }
\label{Fig:PS_NV4}
\end{figure}

Performing the alternating sequence $\pi/2(x)$ pulse - dynamical decoupling - $\pi/2(x)$, which transforms our system to state $\ket{1}$ when $\Phi=0$, effectively exchanges the resulting populations $p_0$ and $p_1$. The respective fluorescence signals of the two measurements are then 
\begin{equation}
\begin{aligned}
&\widetilde{F}_0 = p_0 F_0 + p_1 F_1, \\
&\widetilde{F}_1 = F_0 + F_1 - \widetilde{F}_0 = p_1 F_0 + p_0 F_1,
\end{aligned}
\end{equation}
where $\widetilde{F}_0$ ($\widetilde{F}_1$) is the expected fluorescence after the first (alternative) sequence.  Finally, we obtain the contrast for the power spectrum measurement
\begin{align}
c_{\text{ps}}=\widetilde{F}_0-\widetilde{F}_1=c_\text{max}(p_0-p_1)=c_\text{max}\cos\left(\Phi\right),
\end{align}
where we used that $\cos\left(\Phi\right)=\cos\left(\frac{\Phi}{2}\right)^2-\sin\left(\frac{\Phi}{2}\right)^2$.
Multiple readouts are performed and we average the result, which leads to
\begin{align}
\overline{c}_{\text{ps}}=c_\text{max}\left\langle\cos\left(\Phi\right)\right\rangle.
\end{align}
Assuming that the phase $\Phi$ follows a Gaussian distribution, centred at zero with a variance $\langle \Phi^2\rangle$, we obtain
\begin{align}
\overline{c}_{\text{ps}}=c_\text{max}\exp\left(-\langle\Phi^2\rangle/2\right)\approx c_\text{max}\left(1-\langle\Phi^2\rangle/2\right),
\end{align}
where the last approximation is valid only for small $\langle\Phi^2\rangle$. The variance of the phase $\Phi$ depends on the strength of   the interaction between the NV centre and the nuclear spins in the sensing volume, the filter function of the applied DD sequence, and the interaction time. The expected value of the phase variance in case of statistical polarisation is given by $\langle\Phi^2\rangle=\Phi_{\text{rms}}^2$ (see \ref{Section:SP_modeling}) with
\begin{align}\label{Eq:phi_rms2}
\Phi_{\text{rms}}=\frac{2}{\pi}\gamma_e B_{\text{rms}} N\tau
~\text{sinc}{\left(N(\pi-\omega_{\text{L}}\tau)\right)},
\end{align}
where $\gamma_e$ is the gyromagnetic ratio for the NV electron spin, $B_{\text{rms}}$ is the root-mean-square of the magnetic field of the statistically polarised sample, $N$ is the number of the DD pulses, $\tau$ is their pulse separation, and $\omega_{\text{L}}$ is the Larmor frequency of the sensed spins (in angular frequency units). We note that we assumed that the coherence time of the sensed signal is much longer than the sensing time during the DD sequence, so $\left\langle B(t^\prime)B(t^\prime+t)\right\rangle\approx B_{\text{rms}}^2 \cos{(\omega_{\text{L}}t)}$. If this assumption is not feasible the spectrum of the signal from the nuclear spins can be taken into account \cite{Oviedo2020,Pham2016}. However, this effect is typically not large for our power spectrum experiments, so we will neglect in the present analysis. If the DD sequence is applied on resonance, i.e., $\tau=\pi/\omega_{\text{L}}$, the variance of the accumulated phase takes the form
\begin{align}\label{Eq:phi_rms_res}
\Phi_{\text{rms}}=\frac{2}{\pi}\gamma_e B_{\text{rms}} N\tau.
\end{align}
The formulas for the phase $\Phi_{\text{rms}}$ in Eqs.~\eqref{Eq:phi_rms2} and \eqref{Eq:phi_rms_res} allow us to estimate it by a power spectrum measurement, where we vary $\tau$ and keep $N$ constant (see Fig. \ref{Fig:PS_NV4}) or DD order scan measurement where we keep $\tau=\pi/\omega_{\text{L}}$ and vary the number of DD pulses $N$, respectively.

The formula for the expected contrast of the power spectrum is given by
\begin{align}
\overline{c}_{\text{ps}}=c_\text{max}\exp{(-N\tau/T_2)}\exp\left(-\Phi_{\text{rms}}^2/2\right),
\end{align}
where $T_2$ is the coherence time of the respective DD sequence. We have neglected the very small effect of $T_1$ decay during the usually very short duration of the DD sequence in comparison to the $T_1$ time. We note that DD is not efficient for compensation of high frequency noise and pulse errors might also lead to a loss of the contrast $c_\mathrm{max}$ even without an external signal, i.e., for $\Phi=0$. 

\subsection{Correlation spectroscopy}
We measure the population of the NV centre in the correlation spectroscopy experiments analogously to the power spectrum measurement. The pulse sequence is illustrated in Figure \ref{Fig:pulse_sequences}(a). 
%
%After the second dynamical decoupling sequence the NV centre is in a state $\ket{\phi} = \frac{\ket{0} + \exp{(i\phi)} \ket{1}}{\sqrt{2}}$. 
We again apply two alternating measurements, where the spin state of the NV centre is projected on either $\ket{0}$ or $\ket{1}$ with a $\pi/2$-pulse around the $y$- or $-y$-axis (see Methods). 

In the following we describe in more detail the state of the system at each step of the correlation spectroscopy experiment.
%\begin{gather}
%\pi/2(x)- \text{DD sequence} -  \pi/2(y) \notag\\
%- \text{ waiting time $T$ } -  \notag\\
%\pi/2(x)- \text{DD sequence} -  \pi/2(\pm y),
%\end{gather}
%Specifically, if 
We assume that the system is initially prepared in state $|0\rangle$ by optical pumping. Then, the density matrix after the first correlation spectroscopy sequence $\pi/2(x)- \text{DD sequence} -  \pi/2(y)$ is 
\begin{align}
\rho_{\text{cs,1}}
&=\frac{1}{2}\left[ \begin{array}{cc} 1-\sin\left(\Phi_1\right) & i\cos\left(\Phi_1\right) \\
-i\cos\left(\Phi_1\right)  & 1+\sin\left(\Phi_1\right) \end{array}\right],
\label{eq:rho_corrspec1}
\end{align}
where $\Phi_1$ is the accumulated phase during the first DD sequence and we assumed negligible relaxation during the sequence. The off-diagonal elements become zero due to decoherence during the waiting time $T\gg T_2^{\ast}$ and the density matrix afterwards is  
\begin{align}
\rho_{\text{wait}}
&=\frac{1}{2}\left[ \begin{array}{cc} 1-\sin\left(\Phi_1\right) & 0 \\
0  & 1+\sin\left(\Phi_1\right) \end{array}\right].
\end{align}
Finally, the density matrix after the second correlation spectroscopy sequence $\pi/2(x)- \text{DD sequence} -  \pi/2(y)$ is
\begin{align}
\rho_{\text{cs,2}}
&=\frac{1}{2}\left[ \begin{array}{cc} 1+\sin\left(\Phi_1\right)\sin\left(\Phi_2\right) & -i\sin\left(\Phi_1\right)\cos\left(\Phi_2\right) \\
i\sin\left(\Phi_1\right)\cos\left(\Phi_2\right)  & 1-\sin\left(\Phi_1\right)\sin\left(\Phi_2\right) \end{array}\right]
\end{align}
with the accumulated phase $\Phi_2$ during the second DD sequence and we again assumed negligible relaxation during the sequence. The population of state $|0\rangle$ is then $p_0=\frac{1}{2}\left( 1+\sin\left(\Phi_1\right)\sin\left(\Phi_2\right)\right)$ and the population of state $|1\rangle$ is $p_1=\frac{1}{2}\left( 1-\sin\left(\Phi_1\right)\sin\left(\Phi_2\right)\right)$. The populations of the two states are exchanged with the alternative sequence $\pi/2(x)- \text{DD sequence} -  \pi/2(-y)$. 
%
%The respective fluorescence signals after the first and alternating sequence are then 
%\begin{equation}
%\begin{aligned}
%\widetilde{F}_0 &= p_0 F_0 + p_1 F_1, \\
%\widetilde{F}_1 &= F_0 + F_1 - \widetilde{F}_0 = p_1 F_0 + p_0 F_1.
%\end{aligned}
%\end{equation}
%with $\avg{p_i}$ the (average) populations of the final state after projection with the $\pi/2$-pulse around the $y$-axis. 
%These are determined according to the accumulated phases $\Phi_1$ and $\Phi_2$ during the dynamical decoupling sequences. 
Again, the measurement signal is the difference between them
\begin{align}
c_{\text{cs}} &= \widetilde{F}_0 - \widetilde{F}_1 = c_\mathrm{max} (p_0 - p_1) \notag\\
&= c_\mathrm{max} \sin\left(\Phi_1\right)\sin\left(\Phi_2\right)\approx c_\mathrm{max} (\Phi_1 \Phi_2), 
\end{align}
where the last approximation is valid for small accumulated phases during each DD sequence, which is typically the case. We perform multiple readouts and average the result, which leads to
\begin{align}
\overline{c}_{\text{cs}}=c_\mathrm{max} \avg{\Phi_1 \Phi_2}= c_\mathrm{max}  \Phi_{\text{rms}}^2 \cos{(\omega t)}C(t)
\end{align}
with the envelope $C(t)$ of the correlations (see \ref{Section:SP_modeling} for details). 

\subsection{Qdyne}
In the Qdyne measurements the NV centre's PL is not normalised owing to the fact that the individual readouts are not averaged but stored individually. The photon count rates of our setup were of around $\eta_0 = 0.04$ and $\eta_1 = 0.03$, i.e. in 100 readouts in average four (three) photons are detected if the spin state is $\ket{0}$ ($\ket{1}$). Again, the readout window is optimised to achieve the highest SNR and all other photons during the laser pulse that are not within the readout window are discarded. 

\begin{figure*}[t!]
%\centering
%\includegraphics[trim=0cm 0cm 0cm 0cm, clip=true,width=10cm, angle=0]{CS_scheme.pdf}
\includegraphics[width=0.6\textwidth]{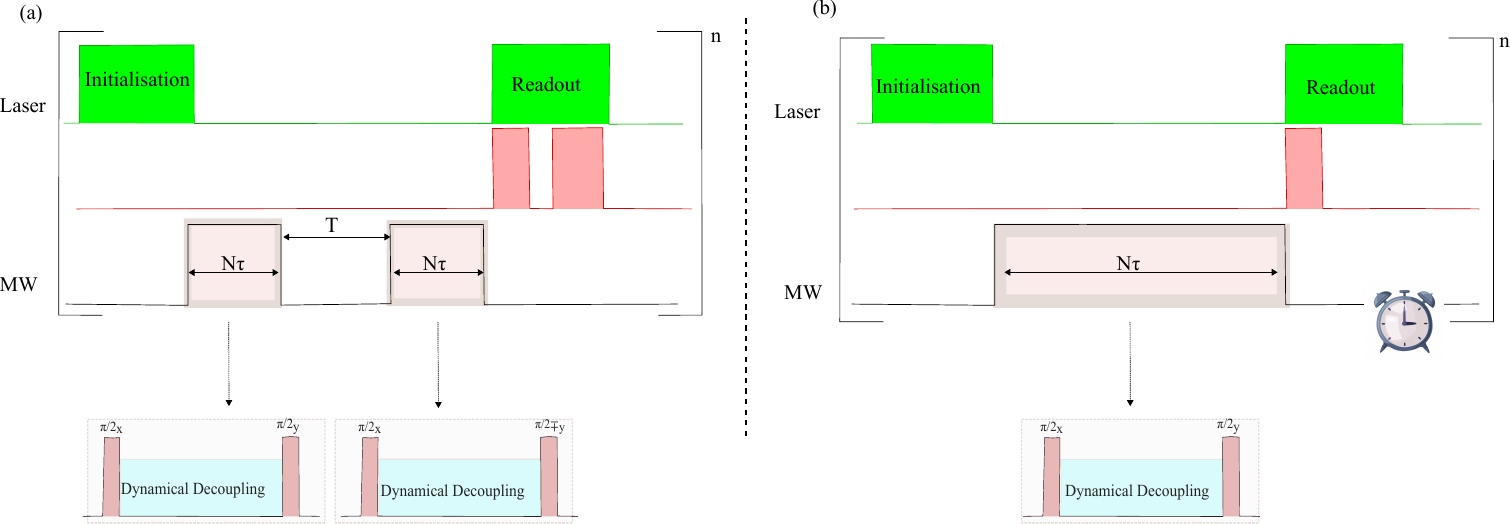}
\caption{Pulse sequences for the used protocols. (a) Correlation spectroscopy pulse sequence. The NV centres are initialised into $m_s= 0$ state using a green laser. %excitation followed by MW manipulation. 
The MW pulse sequence consists of two DD sequences with equal length $N \tau$ where $N$ is the total number of $\pi$-pulses and $\tau = \frac{1}{2f_\mathrm{L}}$ is the spacing between two subsequent $\pi$-pulses where $f_\mathrm{L}$ is the Larmor frequency of the nuclei. The dynamical decoupling sequence is sandwiched between two $\pi/2$-pulses of orthogonal phases. The final spin state is optically measured using the green laser where the signal counts (first red window) are normalised with the reference counts (second red window). The measurement is averaged over $n$ sweeps.(b) Qdyne pulse sequence. The measurement sequence with a single DD block is repeated continuously. The measurement result during each sweep is stored individually and the sequence synchronised before next sweep starts. The photon counts (red window) are not normalised. 
%A spectral filter is generated with the dynamical decoupling sequence where the centre frequency of the filter is determined by the spacing between the $\pi$ pulses, $\tau$ and the width of the filter is inversely proportional to the total duration of the dynamical decoupling sequence, $N \tau$ as in DD power spectrum and correlation spectroscopy sequences. Here, the filter centre frequency is set to $f_{L}$ of the nuclei we intend to detect. The amount of dephasing, $\phi$ of the NV centre during the dynamical decoupling depends on the initial phase of the signal field. The measurement result during each sweep is recorded and synchronised with an external clock before next sweep starts where the phase of the signal is different.
}
\label{Fig:pulse_sequences}
\end{figure*}

The measurement sequence (see Figure \ref{Fig:pulse_sequences}(b)) is similar to the first step of correlation spectroscopy, so the density matrix is the same as in \eqref{eq:rho_corrspec1}
\begin{equation}
\rho_\mathrm{Qd} =\frac{1}{2}
\left[ \begin{array}{cc} 
1-\sin\left(\Phi\right) & i\cos\left(\Phi\right) \\
-i\cos\left(\Phi\right) & 1+\sin\left(\Phi\right) 
\end{array} \right]
\end{equation}
with the accumulated phase $\Phi$ during dynamical decoupling. The population oscillates around $1/2$ for which reason the signal in Eq.~(8) of the main text is measured. 

%%%%%%%%%%%%%%%%%%%%%%%%%%%%%%%%%%%%%%%%%%%%%%%%%%%%%%%%%%%%%%%%%%%%%%%%%%%%%%%%%%%%%%%%%%%%%%%%%%%%%%%%%%%%%%%%%%%%%%%%%%%%%%%
%%%%%%%%%%%%%%%%%%%%%%%%%%%%%%%%%%%%%%%%%%%%%%%%%%%%%%%%%%%%%%%%%%%%%%%%%%%%%%%%%%%%%%%%%%%%%%%%%%%%%%%%%%%%%%%%%%%%%%%%%%%%%%%
\section{Modelling statistical polarisation}\label{Section:SP_modeling}
%%%%%%%%%%%%%%%%%%%%%%%%%%%%%%%%%%%%%%%%%%%%%%%%%%%%%%%%%%%%%%%%%%%%%%%%%%%%%%%%%%%%%%%%%%%%%%%%%%%%%%%%%%%%%%%%%%%%%%%%%%%%%%%
%%%%%%%%%%%%%%%%%%%%%%%%%%%%%%%%%%%%%%%%%%%%%%%%%%%%%%%%%%%%%%%%%%%%%%%%%%%%%%%%%%%%%%%%%%%%%%%%%%%%%%%%%%%%%%%%%%%%%%%%%%%%%%%
%
The signal we observe is due to statistical polarisation in the detection volume of the NV centre(s). We assume that the magnetic field, generated by the sample due to statistical polarisation is given by \cite{Gefen2019}
\begin{align}
B(t)=A\cos{(\omega t)}+D\sin{(\omega t)},
\end{align}
where $A,D\sim N(0,B_{\text{rms}})$ the normal distribution with zero mean and standard deviation $\brms$, so the variance of the magnetic field is $\left\langle B(t)^2\right\rangle=B_{\text{rms}}^2$. When we apply a DD sequence on resonance, i.e., by instantaneous $\pi$ pulses at times $\pi(1+2k)/\omega,\ k=0\dots n$ we can sense only $A\cos{(\omega t)}$. The accumulated phase takes the form \cite{Gefen2019}
\begin{align}
\Phi&=\int_{0}^{N \tau}\gamma_e B(t^\prime)f(t^\prime)d t^\prime\notag\\
&=
\gamma_e\int_{0}^{N \tau} A|\cos{(\omega t^\prime)}|d t^\prime \approx \frac{2}{\pi}\gamma_e A N \tau,
\end{align}
where $f(t)$ is a modulation function due to the DD pulses and $N \tau$ is the duration of the DD sequence with $N$ the number of pulses and $\tau=\pi/\omega$ the time between the centres of the pulses. Since $A\sim N(0,B_{\text{rms}})$, we obtain
\begin{align}\label{Eq:var_phi_stat}
\langle\Phi\rangle&=0,\notag\\
\langle\Phi^2\rangle&=\left\langle\left(\frac{2}{\pi}\gamma_e A N \tau\right)^2\right\rangle=\left(\frac{2}{\pi}\gamma_e B_{\text{rms}} N \tau\right)^2\equiv \Phi_{\text{rms}}^2.
\end{align}
%
% We note that an equivalent way to model the noise is by assuming
% \begin{align}
% B(t)=g\cos{(\omega t+\phi)},
% \end{align}
% where $g = \sqrt{A^2+D^2}\sim N(0,B_{\text{rms}}\sqrt{2})$, $\phi=\arctan{\left(D/A\right)}\sim U(0,2\pi)$. Then, the accumulated phase with DD on resonance takes the form
% \begin{align}
% \Phi&=\int_{0}^{t}\gamma_eB(t^\prime)f(t^\prime)\d t^\prime\notag\\
% &=\gamma_e\int_{0}^{t}
% g\left(\cos{(\omega t^\prime)}\cos{(\phi)}+\sin{(\omega t^\prime)}\sin{(\phi)}\right)
%  f(t^\prime)\d t^\prime\notag\\
% &=
% \gamma_e\int_{0}^{t}g\cos{(\phi)}|\cos{(\omega t^\prime)}|\d t^\prime
% \approx \frac{2}{\pi}\gamma_e g \cos{(\phi)} t.
% \end{align}
% Then, the expected value of the accumulated phase and its variance take the form
% \begin{align}
% \langle\Phi\rangle&=0,\\
% \langle\Phi^2\rangle&=\left\langle\left(\frac{2}{\pi}\gamma_e g \cos{(\phi)} t\right)^2\right\rangle \notag\\
% &=\left(\frac{2}{\pi} \gamma_e t\right)^2 \left\langle g^2\right\rangle \frac{1}{2\pi}\int_{0}^{2\pi}\cos{(\phi)}^2 d \phi\notag\\
% &=\left(\frac{2}{\pi} \gamma_e t\right)^2 2 B_{\text{rms}}^2 \frac{1}{2}=\left(\frac{2}{\pi}\gamma_e B_{\text{rms}} t\right)^2= \Phi_{\text{rms}}^2\notag,
% \end{align}
% which the same as the value obtained in Eq. \eqref{Eq:var_phi_stat}.

Next, we calculate the expected value of $\langle\Phi_1\Phi_2\rangle$ of two phases $\Phi_1$ and $\Phi_2$, accumulated by two DD sequences, starting at times $t_{1}$ and $t_2=t_{1}+t$, where $t$ %$t=N\tau+T$ 
is the difference between the starting times of the two DD sequences. %with $N\tau$ - the total duration of the DD sequence and $T$ is the spacing between the two DD sequences e.g., as in the case of correlation spectroscopy. 
First, it is useful to state that we assume now that $A$ and $D$ can be time dependent
\begin{align}
B(t)&=A(t)\cos{(\omega t)}+D(t)\sin{(\omega t)}
\end{align}
as the time separation $t$ between the starting times of the two DD sequences can be much longer than the diffusion time of the sample, which results in a change in $A$ and $D$ between the two sequences.
However, we also assume for simplicity that their values do not change during a DD sequence ($N\tau \ll T_D$), thus they depend only on the starting time of the respective sequence. Finally, we also assume $\langle A(t_1)A(t_2)\rangle=\langle D(t_1)D(t_2)\rangle=B_{\text{rms}}^2 C(t_2-t_1)$, $\langle A(t_k)D(t_l)\rangle=\langle D(t_k)A(t_l)\rangle=0, \ k,l=1,2$. The correlation function $C(t_2-t_1)=C(t)$ of the magnetic field envelope can have a different shape that we probe, e.g., exponential decay \cite{Pham2016} or a power-law decay at long times \cite{Cohen2020}. We obtain
\begin{align}
\left\langle B(t_1)B(t_2) \right\rangle&=\left\langle A(t_1)A(t_2)\cos{(\omega t_1)}
\cos{(\omega (t_1+t))}\right\rangle\notag\\
&+\left\langle D(t_1)D(t_2)\sin{(\omega t_1)}
\sin{(\omega (t_1+t))}\right\rangle\notag \notag\\
&= B_{\text{rms}}^2 \cos{(\omega t)}C(t),
\end{align}
where we used $\langle A(t_k)D(t_l)\rangle=0$ in the first equality.
The accumulated phases during the first DD sequence then take the form
\begin{align}
\Phi_1&=\int_{0}^{N \tau}\gamma_e B(t^\prime)f(t^\prime)d t^\prime\\
&=
\gamma_e A(0)\int_{0}^{N \tau}|\cos{(\omega t^\prime)}|d t^\prime \approx \frac{2}{\pi}\gamma_eA(0) N \tau. \notag
\end{align}
We took $t_1=0$ for simplicity of presentation and without loss of generality. For the second DD sequence 
\begin{align}
\Phi_2&=\int_{t}^{t+N\tau}\gamma_e B(t^\prime)f(t^\prime)d t^\prime\notag\\
&=\gamma_e\left(A(t)\cos{\left(\omega t\right)}+D(t)\sin{\left(\omega t\right)}\right)\int_{t}^{t+N\tau}|\cos{(\omega t^\prime)}|d t^\prime \notag\\
&\approx \frac{2}{\pi}\gamma_e\left(A(t)\cos{\left(\omega t\right)}+D(t)\sin{\left(\omega t\right)}\right) N\tau,
\end{align}
where $N \tau$ is again the duration of the DD sequence. Then, the expected value of $\langle\Phi_1\Phi_2\rangle$ takes the form
\begin{align}
\left\langle \Phi_1 \Phi_2 \right\rangle&\approx \left(\frac{2}{\pi} \gamma_e N \tau\right)^2\left\langle  A(0)A(t)\right\rangle\cos{\left(\omega t\right)}\notag\\
&= \left(\frac{2}{\pi} \gamma_e B_{\text{rms}} N \tau\right)^2 \cos{(\omega t)}C(t)\notag\\
&= \Phi_{\text{rms}}^2 \cos{(\omega t)}C(t),
\end{align}
where we used $\langle A(0)D(t)\rangle=0$ in the first equality.

\section{Power spectrum shape analysis}\label{Cohenianderivation}
In the following, we derive the asymptotic behaviour of the power spectrum of the magnetic noise induced on the NV centre by diffusing particles in three spatial dimensions. This derivation is based on the supplementary material of \cite{Cohen2020}.

We assume that the nuclei obey the diffusion equation. The diffusion is therefore described by the equation
\begin{equation}\label{Asymp1}
    \left(\partial_t-D\nabla^2\right)P\left(\vec{r},\vec{r_0},t\right)=\delta\left(\vec{r}-\vec{r}_0\right),
\end{equation}
where $P\left(\vec{r},\vec{r_0},t\right)$ is the conditional probability that a particle is found at position $\vec{r}$ at time $t$ if it was initially at position $\vec{r_0}$ at time $t=0$, and $\delta$ is the Dirac delta function.
The power spectrum is the Fourier transform of the correlation
\begin{equation}\label{Asymp2}
    C(t)=\left<B(t)B(0)\right>\propto\int\frac{d^3 r}{r^3}\int\frac{d^3 r_0}{r_0^3}P\left(\vec{r},\vec{r_0},t\right).
\end{equation}
Since the time dependence enters only through the conditional probability $P\left(\vec{r},\vec{r_0},t\right)$ % propagator, 
taking the Fourier transform of \eqref{Asymp2} yields
\begin{equation}\label{Asymp3}
    S\left(\omega\right)\propto   \textrm{Re}\left\{\int\frac{d^3 r}{r^3}\int\frac{d^3 r_0}{r_0^3}P_\omega\left(\vec{r},\vec{r_0}\right)\right\},
\end{equation}
where 
\begin{equation}\label{Asymp4}
    \left(\textrm{i}\omega-\frac{D}{2\pi}\nabla^2\right)P_\omega\left(\vec{r},\vec{r_0}\right)=\delta\left(\vec{r}-\vec{r}_0\right).
\end{equation}
At $\omega=0$ the solution to Eq. \eqref{Asymp4} is equivalent to that of the electrostatic potential of a point charge
\begin{equation}\label{Asymp5}
        P_{\omega=0}\left(\vec{r},\vec{r_0}\right)=\frac{1}{2D}\frac{1}{\left|\vec{r}-\vec{r}_0\right|}.
\end{equation}
Substituting the conditional probability %propagator 
\eqref{Asymp5} into Eq. \eqref{Asymp3} leads to 
\begin{equation}\label{Asymp6}
    S\left(\omega=0\right)\propto   \frac{1}{D}\int_{r>d}\frac{d^3 r}{r^3}\int_{r_0>d}\frac{d^3 r_0}{r_0^3}\frac{1}{\left|\vec{r}-\vec{r}_0\right|},
\end{equation}
where the lower bound for the integration is determined by the depth $d$ of the NV centre. The only length scale in the problem is the NV's depth, therefore, by dimensional analysis of Eq. \eqref{Asymp6},
\begin{equation}\label{Asymp7}
 S\left(\omega=0\right)\propto\frac{C_1}{dD},
\end{equation}
where $C_1$ is a non-universal constant that depends on the geometry of the problem.

The first correction in frequencies to the spectrum Eq.~\eqref{Asymp7}, in the limit $\omega\ll\omega_D$, where the diffusion time $T_D$ is the characteristic time to diffuse at a distance $d$, i.e., $d=\sqrt{D T_D}$ with $\omega_D =2\pi/T_D=\frac{2\pi D}{d^2}$ the characteristic angular frequency of diffusion noise, can be accounted for by considering the finite diffusion length scale $l_\omega=\sqrt{2\pi D/\omega}\gg d$. 
This creates an effective cut-off to the interaction, such that Eq. \eqref{Asymp6} is corrected to
%\begin{multline}\label{Asymp8}
%    S\left(\omega\ll\omega_D\right) \propto   \frac{1}{D}\int_{r>l_\omega}\frac{d^3 r}{r^3}\int_{r_0>l_\omega}\frac{d^3 r_0}{r_0^3}\frac{1}{\left|\vec{r}-\vec{r}_0\right|}\propto \\\frac{1}{dD}\left(C_1-C_2\sqrt{\frac{\omega}{\omega_D}}\right), 
%\end{multline}
\begin{multline}\label{Asymp8}
    S\left(\omega\ll\omega_D\right) \propto \frac{1}{D}\int_{d}^{l_\omega}\frac{d^3 r}{r^3}\int_{d}^{l_\omega}\frac{d^3 r_0}{r_0^3}\frac{1}{\left|\vec{r}-\vec{r}_0\right|}\propto
    \\\frac{C_1}{dD} - \frac{1}{D}\int_{r>l_\omega}\frac{d^3 r}{r^3}\int_{r_0>l_\omega}\frac{d^3 r_0}{r_0^3}\frac{1}{\left|\vec{r}-\vec{r}_0\right|}\propto \\\frac{1}{dD}\left(C_1-C_2\sqrt{\frac{\omega}{\omega_D}}\right), 
\end{multline}
where again $C_2$ is a non-universal constant. This reproduces the sharp-peaked behaviour around the Fourier peak of the power spectrum shown in Fig.~2(d) on the main text.

\begin{figure*}[t!]
\includegraphics[width=\textwidth]{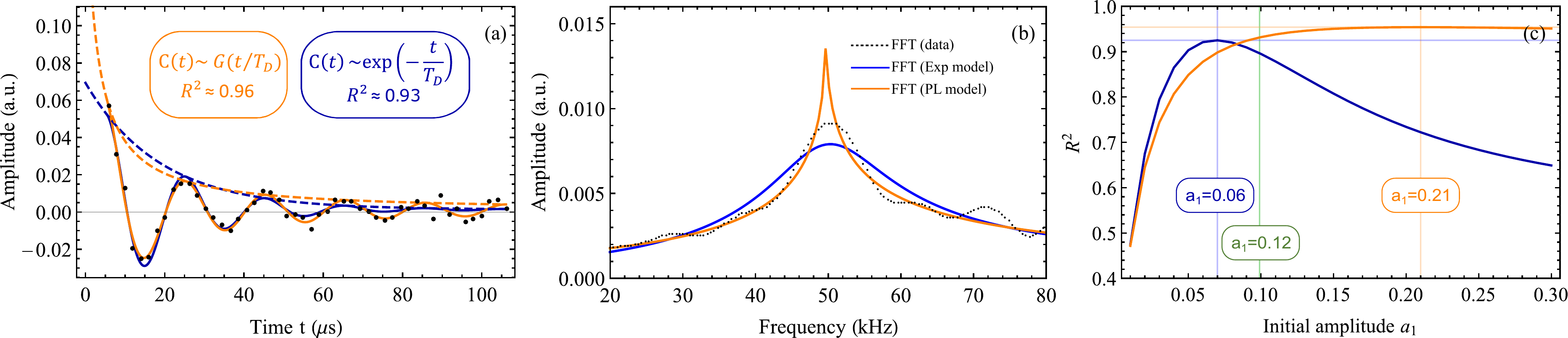}
\caption{Experimental data from correlation spectroscopy measurement with a single NV centre. (a) Signal of the correlation spectroscopy measurement vs. time $t$ between the beginnings of the two dynamical decoupling sequences in correlation spectroscopy (black dots) and fits of the exponential (blue line) and power-law models (orange line) with their goodness of fit. The power law model shows a better fit to the data, especially at long times. There is also a significant difference in the estimate of the initial amplitude of the oscillation $a_1$. (b) FFT
of the experimental data, FFT from the fitted data to the
exponential and the power-law decay models (in (a)) is shown
for illustrative purposes. The results show that the power law
model allows for a more precise frequency estimation. (c) Optimum $R^2$ for different fixed values of the initial amplitude $a_1$ of the correlation spectroscopy signal for the single NV experiment. The blue (orange) curves shows the best fits of the data to the function $a_0+a_1\cos{(\omega t+\phi)}C(t)$ for the exponential (power-law) decay model while the green lines show the estimated values of the initial amplitude $a_1=0.12$, obtained from independent power spectrum measurements. The peak value of the $R^2$ curve is higher for the power-law model in comparison to the exponential decay model. In addition, the power-law model is substantially better when the initial amplitude is fixed to the estimate from the independent power spectrum measurement, i.e., comparing the $R^2$ of the two models for $a_1$ on the green lines. The fits with fixed $a_1$ and their FFT are shown in Figure~2 of the main text. 
}
\label{Fig:a1_fits}
\end{figure*}

The correction can be understood intuitively by considering that sensing an oscillating signal at a small angular frequency $\omega$ (or two signals with an angular frequency difference $\omega$) typically requires some non-zero correlation $C(t)$ at time $t=t_{\omega}=2\pi/\omega$, so we could detect at least one oscillation cycle of the signal. In the regime $t_{\omega}\gg t_D$, or equivalently $\omega\ll\omega_D$, this leads to a minimum time $t_{\omega}$ where any oscillating component  of the auto-correlation function at $\omega$ after this time contributes to a lower value of $S(\omega)$ in comparison to $S(\omega=0)$ and thus a narrower linewidth. 
Spatially, this means that there is an effective cut-off length $l_\omega=\sqrt{2\pi D/\omega}\gg d$, where the spectral density at $\omega$, i.e., $S(\omega)$ is smaller than $S(\omega=0)$ due to an oscillating signal at $\omega$ coming from nuclei which diffuse more than the cut-off interaction length $l_\omega$.

In the other limit of large frequencies, e.g. $\omega\gg\omega_D$, which is equivalent to $l_\omega\ll d$, we approximate $D\nabla^2\approx-D/d^2\sim -\omega_D$ as the small correction to the large frequency $\omega$. Eq. \eqref{Asymp4} then takes the form
\begin{equation}\label{Asymp9}
    P_{\omega\gg\omega_D}\propto d^{-3}\frac{\delta\left(\vec{r}-\vec{r_0}\right)}{-i\omega+\omega_D}.
\end{equation}
Hence, the spectrum, Eq. \eqref{Asymp3}, approaches
\begin{equation}\label{Asymp10}
    S(\omega\gg\omega_D)\propto\text{Re}\left\{  d^{-3}\frac{\delta\left(\vec{r}-\vec{r_0}\right)}{-i\omega+\omega_D}\right\}\approx S\left(\omega=0\right) \left(\frac{\omega_D}{\omega}\right)^2.
\end{equation}
Eq. \eqref{Asymp10} recovers the well known Lorentzian behaviour of the spectrum. 

Note that whether one considers the angular dependence and the geometry changes the prefactors but not the overall behaviour of the correlation function or the spectrum shape.

\section{Additional experimental details}\label{Section:Experimental_details}
%%%%%%%%%%%%%%%%%%%%%%%%%%%%%%%%%%%%%%%%%%%%%%%%%%%%%%%%%%%%%%%%%%%%%%%%%%%%%%%%%%%%%%%%%%%%%%%%%%%%%%%%%%%%%%%%%%%%%%%%%%%%%%%
%%%%%%%%%%%%%%%%%%%%%%%%%%%%%%%%%%%%%%%%%%%%%%%%%%%%%%%%%%%%%%%%%%%%%%%%%%%%%%%%%%%%%%%%%%%%%%%%%%%%%%%%%%%%%%%%%%%%%%%%%%%%%%%

We measure the auto-correlation function of the detected signal by three different approaches: correlation spectroscopy with a single shallow NV centre, correlation spectroscopy with an ensemble of NV centres, and Qdyne with single NV centres. Next, we describe more details on the experimental setup and measurements results for each of these.

\subsection{Experimental setup for single NV centre experiments}

%The experiments with single NV centres are carried out on shallow NV centres of an isotopically enriched (99.99$\%$ $^{12}$C) diamond sample grown by chemical vapour deposition. The diamond substrate%, purchased from Element Six Inc., 
 %with natural abundance of $^{13}$C was equipped with a $99.999\%$ $^{12}$C enriched homoepitaxially grown diamond film with a thickness of about 150\,nm in a home-built plasma enhanced chemical vapour deposition growth reactor\cite{osterkamp2019engineering,silva2010microwave}. Isolated, shallow NV centres were then created by ion implantation with $^{15}$N$^+$ at a dose of $5 \times 10^{15}\,\mathrm{N^+/cm^2}$ using an acceleration energy of 2\,keV (correlation spectroscopy) and 2.5\,keV (Qdyne). To heal the resulting radiation damage, mobilise vacancies and eventually create the desired NV centres, the diamond is annealed in a home-built UHV furnace at $1000\,^\circ\mathrm{C}$ for 3 hours, while ensuring extremely low process pressures $< 1 \times 10^{-7}$\,mbar []. After annealing, the diamond is boiled in a 1:1:1 mixture of sulfuric, perchloric and nitric acid at $200 \,^\circ\mathrm{C}$ for 30 minutes to remove any (graphitic) residues from the surface\cite{lang2020long,findler2020indirect}.% 

Experiments for the correlation spectroscopy (CS) and Qdyne measurements have been done on two different but conceptually equivalent setups. The shallow NV centres are addressed via a fluorescent confocal scan done on a home-built confocal microscope. The NV centres can be initialised and read out using a 532\,nm (CS) and 517\,nm (Qdyne) laser pulse of about 5000\,ns (CS) and 1000\,ns (Qdyne) duration generated by a CW laser (Laser Quantum Gem532, Toptica iBeam smart 515) and chopped by an acousto-optic modulator (CS, Crystal Technology 3200-146) or by the pulsed-laser output itself (Qdyne). 
The pulse sequences containing both, trigger signals for laser output pulses and microwave waveforms, are sampled on an AWG (Tektronix AWG70000A, Keysight M8195A), amplified (Amplifier Research 60S1G4, Amplifier Research 30S1G6) and applied to the NV centre through a copper wire of $20\,\mathrm{\mu m}$ diameter strapped across the diamond sample. A single photon counting module (SPCM, Excelitas SPCM-AQRH-4X-TR) is used for detecting the photoluminescence (PL). The SPCM  generates a TTL output whenever it detects a photon.  These signals are recorded with a multiple-event-time digitiser (FAST ComTec P7887, FAST ComTec MCS6A) with time stamp with a timing resolution of 200\,ps. We use the Qudi software suite to orchestrate and control the experiment hardware \cite{Qudi}. 

\begin{figure*}[t!]
\includegraphics[width=\textwidth]{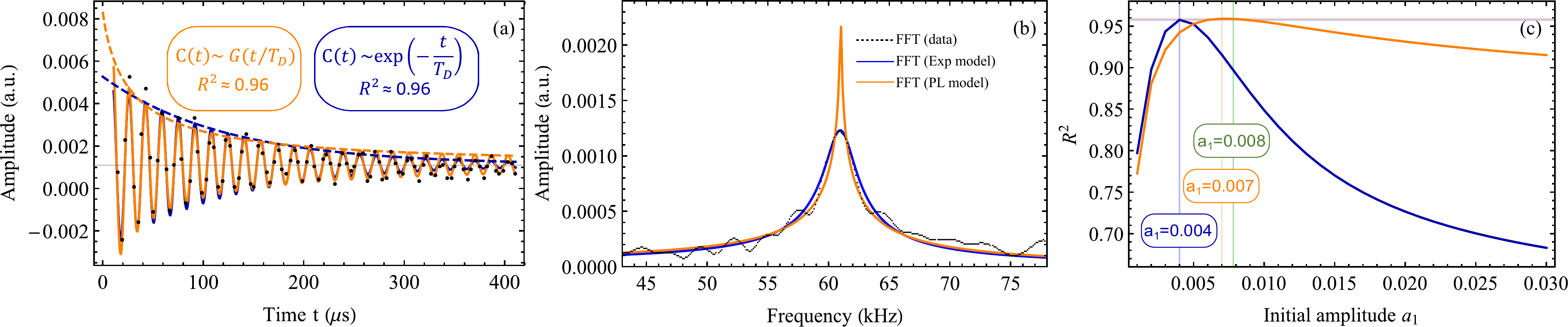}
\caption{Experimental data from correlation spectroscopy measurement with an ensemble of NV centres. (a) Signal of the correlation spectroscopy measurement vs. time $t$ between the beginnings of the two dynamical decoupling sequences in correlation spectroscopy (black dots) and fits of the exponential (blue line) and power-law models (orange line) with their goodness of fit. The power law model shows a similar overall fit to the data as the exponential model, nonetheless fitting the data better at long times. There is a significant difference in the estimate of the initial amplitude of the oscillation $a_1$. (b) FFT
of the experimental data, FFT from the fitted data to the
exponential and the power-law decay models (in (a)) is shown
for illustrative purposes. The results show that the power law
model allows for a more precise frequency estimation. (c) Optimum $R^2$ for different fixed values of the initial amplitude $a_1$ of the correlation spectroscopy signal for the single NV experiment. The blue (orange) curves shows the best fits of the data to the function $a_0+a_1\cos{(\omega t+\phi)}C(t)$ for the exponential (power-law) decay model while the green lines show the estimated values of the initial amplitude $a_1=0.079$, obtained from independent power spectrum measurements. The peak value of the $R^2$ curve is higher for the power-law model in comparison to the exponential decay model. In addition, the power-law model is substantially better when the initial amplitude is fixed to the estimate from the independent power spectrum measurement, i.e., comparing the $R^2$ of the two models for $a_1$ on the green lines. The fits with maximum $R^2$ for either model are shown in (a) in this figure and the fits for fixed $a_1=0.079$ is shown in the Fig.~3(a) in the main text. 
}
\label{Fig:a1_fits_nvision}
\end{figure*}

% \begin{figure*}[t!]
% \includegraphics[width=0.8\textwidth]{supp_figs_a1_NV4_nvision.pdf}
% %\includegraphics[trim=0.9cm 0.4cm 0.5cm 0.4cm, clip=true,totalheight=0.4\textheight,width=20cm, angle=0]{DD_spectrum.pdf}
% \caption{Optimum $R^2$ for different fixed values of the initial amplitude $a_1$ of the correlation spectroscopy signal for (a) the single NV and (b) for the NV ensemble experiments. The blue (red) curve shows the best fits of the data to the function $a_0+a_1\cos{(\omega t)}C(t)$ for the exponential (power-law) decay model while the green lines show the estimated values of the initial amplitude $a_1=0.12$ and $a_1=0.008$, for the single NV and ensemble experiments, respectively, obtained from independent power spectrum measurements. In both experiments the peak value of the $R^2$ curve is higher for the power-law model in comparison to the exponential decay model. In addition, the power-law model is substantially better when the initial amplitude is fixed to the estimate from the independent power spectrum measurement, i.e., comparing the $R^2$ of the two models for $a_1$ on the green lines. The fits with maximum $R^2$ for either model are shown in Figures~\ref{Fig:CS_NV4_v3}(a) and \ref{Fig:CS_NVision}(a) in the main text and with fixed $a_1$ in the Figure subplots (b). 
% }
% \label{Fig:a1_fits}
% \end{figure*}

\subsection{Correlation spectroscopy with a single NV centre}\label{Appendix:CS_single_NV}

%All measurements are performed on a shallow nv called `$NV_4$' of depth $\approx 2.9$ nm.
%For the correlation spectroscopy measurements a single NV centre located at a depth $\approx 2.9$ nm below the surface is used. In the experiments a bias field of $\approx 450$ G is applied using a permanent magnet along the NV axis to lift the $ms = \pm 1$ degeneracy, so the NV centre can be treated as a two-level system by only considering $m_s = 0$ and $m_s = -1$ states. 
The NV centre for the correlation spectroscopy measurements is located $\approx2.9$\,nm below the diamond surface. The depth was estimated by measuring the power spectrum of the hydrogen nuclei in the immersion oil with dynamical decoupling (see Fig.~\ref{Fig:PS_NV4}) \cite{Viola1999}. We used the Knill dynamical decoupling (KDD4) pulse sequence for our experiments \cite{KDDref,KDDref2,CasanovaPRA2015,GenovPRL2017}. KDD has five $\pi$ pulses with phases 
%\begin{equation}
$(\chi+\pi/6,\chi,\chi+\pi/2,\chi,\chi+\pi/6)$ 
%\end{equation}
\cite{KDDref,KDDref2,CasanovaPRA2015,GenovPRL2017}.
It is based on a composite pulse, first proposed by Tycko and Pines \cite{TyckoCPL1984}, which was recently shown to be one example from a set of universal composite pulses for population inversion \cite{GenovPRL2014}. We obtain KDD4 by nesting KDD in the XY4 sequence \cite{KDDref2,CasanovaPRA2015,GenovPRL2017}, i.e., the phase $\chi$ takes values $(0,\pi/2,0,\pi/2)$, resulting in a sequence of twenty pulses. We repeat it two times, which we label as KDD4-2, and vary the pulse time separation \cite{KDDref,KDDref2,CasanovaPRA2015,GenovPRL2017}. %with interaction time of $20.99 \mu s$ for the power spectrum measurement. 
The power spectrum measurement allows estimation of $B_{\text{rms}}$, so we can use it as an input parameter for the data analysis of the correlation spectroscopy measurements in Fig.~2. 
The $T_2$ time with the same DD sequence is measured $110\,\mu s$ by varying the DD order where the spacing between the two adjacent $\pi$ pulses was set off-resonant to $1/(f_{\text{L}} \times 1.3)$. The $T_2^*$ is measured to $200$\,ns using Ramsey sequence. We note that the translation diffusion correlation time for the nuclear spins %to diffuse out of the detection volume 
is given by $T_D = d^2/D_\text{oil}$, where $d$ is the depth of the NV centre and $D_\text{oil} = 5 \times 10^{-13}\,\mathrm{m^2/s}$ is the diffusion coefficient of the immersion oil. The spin lattice relaxation time, $T_1$ of the NV centre is $1.11$\,ms. 

Figure~\ref{Fig:a1_fits} includes additional data analysis of the experimental results from the single NV centre experiment. 
Figure~\ref{Fig:a1_fits}(a) shows that the power-law decay model has a better fit to the data with an $R^2\approx 0.96$ in comparison to $R^2\approx 0.93$ for the exponential model. Its goodness of fit is better especially at long times, e.g., between $50-100\,\mu$s, which correspond to more than three times the expected diffusion time $T_D\approx 17\,\mu$s. In contrast to Fig.~2 in the main text, the initial contrast is considered here a free parameter. We note that the two models give almost a factor of three difference in its estimate (see Fig.~\ref{Fig:a1_fits}(c)). Figure~\ref{Fig:a1_fits}(b)
shows the FFT
of the experimental data, and the FFT from the fitted data to the
exponential and the power-law decay models (in (a)) for illustrative purposes. % It is shown for illustrative purposes.

Figure~\ref{Fig:a1_fits}(c) shows the estimation of the initial amplitude of the auto-correlation function from the correlation spectroscopy data of the single NV centre. The data analysis shows that the overall fit of the power-law model is better, i.e., the peak value of the $R^2$ curve is higher, in comparison to the exponential decay model. However, the optimal values of the initial correlation amplitude $a_1$ for the two models differs substantially. Because $a_1 \propto \frms \propto \brms$ we can estimate it independently with a power spectrum measurement and use the estimate ($a_1=0.12$) as a fixed parameter in the fitting procedure. The result also shows that the power law decay model performs especially well at long times (see Fig. 2 in the main text).

%%%Experimental description to be added.

%\subsubsection{Experimental results}%
%Description of the results in Fig. \ref{Fig:CS_NV4} to be added.

%\subsubsection{Experimental Setup}

%\subsubsection{NV centre characteristics}% NV4

%\begin{figure}[t!]
%\includegraphics[width=0.96\columnwidth]{all_figs_NV4_freq.pdf}
%\includegraphics[trim=0.9cm 0.4cm 0.5cm 0.4cm, clip=true,totalheight=0.4\textheight,width=20cm, angle=0]{DD_spectrum.pdf}
%\caption{Analysis of experimental data from correlation spectroscopy measurements with a single NV centre. Goodness of fit for (a) the standard exponential decay model and (b) the power law model. The power law model shows a better fit to the data, especially at long times. There is also significant variation in the estimate of the initial amplitude of the oscillation $a_1$. (c) Optimum adjusted $R^2$ for different fixed values of the undersampling frequency. Using the power law decay model allow us to obtain both a higher peak adjusted $R^2$ and a narrower estimate than using the exponential model. (d) Optimum adjusted $R^2$ for different fixed values of $a_1$. Goodness of fit fixed $a_1=0.12$, estimated from a power spectrum measurement, obtained by KDD4-2 sequence, for (e) exponential decay model and (f) the power law model. %The experimental data points are fitted with a Lorentzian fit (blue line).
%}
%\label{Fig:CS_NV4}
%\end{figure}

\subsection{Correlation spectroscopy with an NV ensemble}\label{Appendix:CS_ensemble_NV}

The NV ensemble is optically initialised and read out via a self-made widefield setup. In order to increase the collection efficiency, a solid immersion lens with a diameter of \SI{6}{\milli \metre} is placed between the diamond and the objective lens (NA: 0.63, effective focal length: \SI{20}{\milli \metre}, Edmund Optics 85-301). The NV centres are excited by a diode-pumped solid-state laser at \SI{532}{\nano \metre}. An acousto-optical modulator (AOM) (Crystal Technology 3200-146) is used to generate laser pulses. The fluorescence is recorded by a silicon photomultiplier (Excelitas LynX-A-33-A50-T1-A) and measured by a digital oscilloscope (Adlink PXIe-9834). To eliminate back scattered light from the excitation beam, a bandpass filter is placed in front of the detector. Microwave (MW) pulses are first generated by an arbitrary waveform generator (Keysight M3202A) and then amplified. The MW pulses are delivered to the NV sensor through a copper coil placed on top of the diamond. The magnetic field of about 920\,G is generated by an electromagnet and the magnetic field direction is aligned to one of the NV symmetry axis.

Figure~\ref{Fig:a1_fits_nvision} includes additional data analysis of the experimental results from the NV centre ensemble experiment. 
Figure~\ref{Fig:a1_fits_nvision}(a) shows that the power-law decay model has a similar fit to the data with an $R^2\approx 0.96$. Its goodness of fit is better especially at long times, e.g., between $200-400\,\mu$s, which correspond to more than three times the expected diffusion time. % $T_D\approx 17\,\mu$s. 
In contrast to Fig.~3 in the main text, the initial contrast is considered a free parameter, which is the reason behind the exponential model showing very similar fit at long times as the power-law model. Since the aim is to demonstrate the power-law model, our sampling is optimised towards improving the signal-to-noise ratio for long times, where the difference between models is more apparent. But this means that the sampled data does not allow us to efficiently estimate the signal amplitude $a_1$ at very short times (there are very few data points and the $a_1$ confidence interval is quite large). As a result, the fit of unconstrained exponential model tends to underestimate $a_1$, allowing for an artificially increased signal at long times, similar to the power-law model. We note that the two models give almost a factor of two difference in its estimate (see Fig.~\ref{Fig:a1_fits}(c)). Figure~\ref{Fig:a1_fits_nvision}(b) shows the FFT of the experimental data and the FFT from the fitted data to the exponential and the power-law decay models (in (a)) for illustrative purposes. 

Figure~\ref{Fig:a1_fits}(c) includes additional results for the estimation of the initial amplitude of the auto-correlation function from the correlation spectroscopy data of the ensemble measurements. Similarly to the single NV experiments, the data analysis shows that the overall fit of the power-law model is better, i.e., the peak value of the $R^2$ curve is higher, in comparison to the exponential decay model. The optimal values of the initial correlation amplitude $a_1$ for the two models differs substantially, similarly to the single NV experiment. Because $a_1 \propto \frms \propto \brms$ we can estimate it independently with a power spectrum measurement and use the estimate ($a_1\approx 0.008$) as a fixed parameter in the fitting procedure. The result shows that the power law decay model performs especially well at long times (see Fig.~3 in the main text).

%\subsubsection{Chearacteristics of the NV ensemble}% NVision data

%%%Experimental description to be added.

%\subsubsection{Experimental results}%

%Description of the results in Fig. \ref{Fig:CS_NV_ensemble} to be added.

\subsection{Qdyne with single NV centres}\label{Appendix:Qdyne}

\subsubsection{Experimental description}

\begin{figure}[t!]
\includegraphics[width=\columnwidth]{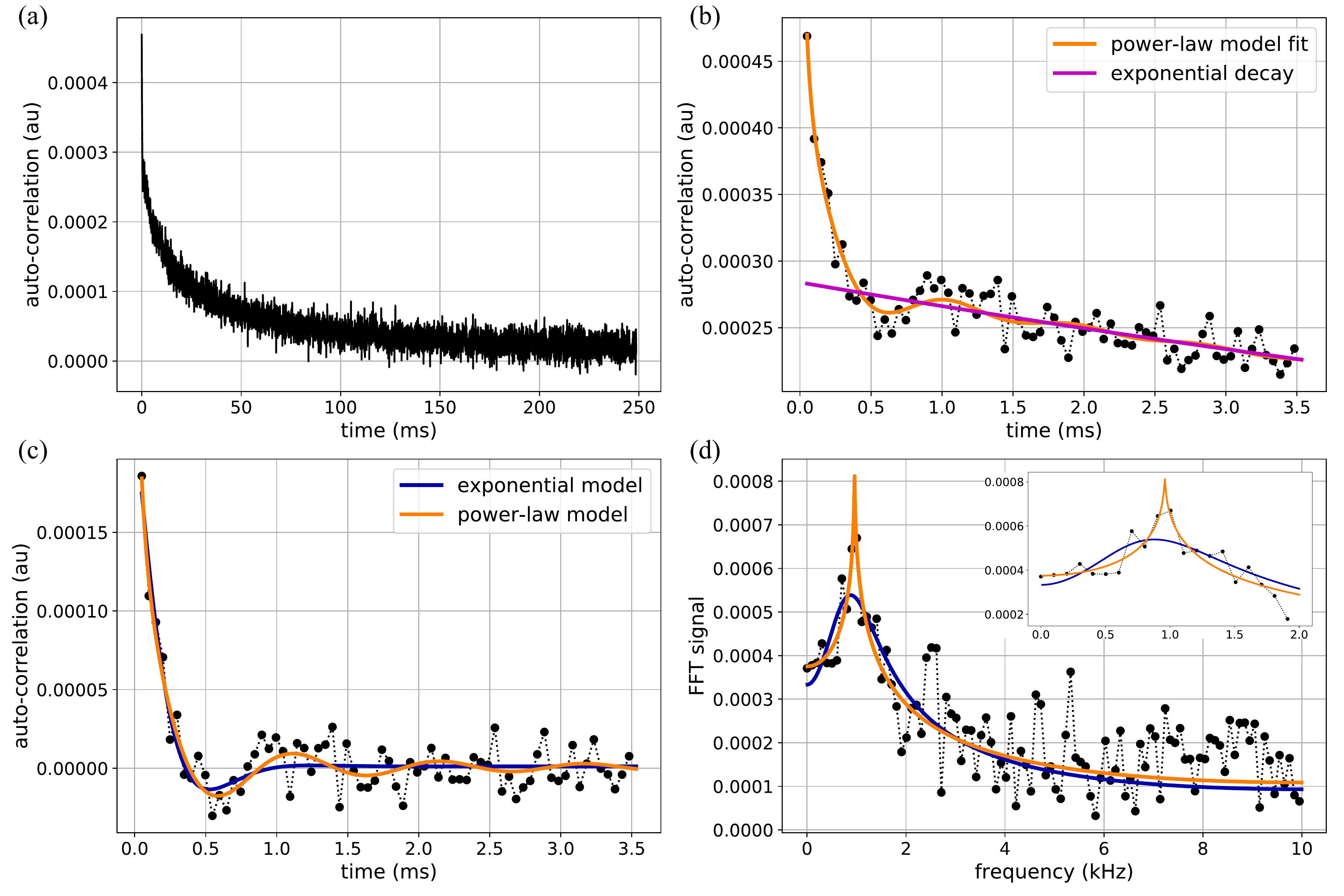}
\caption{Auto-correlation calculation with the Qdyne data. Correlations show exponential decay at the millisecond scale (a). That is removed by fitting an exponential together with the model function (b). In (c) the fit of the exponential and power-law models are shown for the corrected data. FFT of the data and the fitted models for exponential (power-law) models in blue (orange) in (d) with inset showing the peak region.}
\label{Fig:autocorr_calculation}
\end{figure}

For the Qdyne measurements different NV centres located a few nanometers between 8 and 15\,nm below the surface of the diamond are used. %Measurements are carried out employing a confocal microscope with 515\,nm excitation laser which allows initialisation and readout of the NV spin-state. Full NV manipulation is achieved through a copper wire that delivers microwave radiation of choice. Collection of photons is performed via an avalanche photodiode that permits readout of the spin-state with $\approx 25\%$ contrast. 
%In contrast to the correlation spectroscopy measurements where the detected readout photons are averaged for the measurement sequence, in the Qdyne approach the arrival time of each photons is recorded with 200\,ps resolution by the multiple-event-time digitiser and stored individually to computer memory. 
The auto-correlation of the measured time trace is calculated using the \textsl{statstool} package in Python. It typically shows a decay on a time scale of a few ten milliseconds (Fig.~\ref{Fig:autocorr_calculation}(a)), which probably results from fluorescing particles at low concentration that occasionally diffuse through the laser beam bath. To account for this decrease an exponential decay is added to the model function and the data is corrected by subtracting the exponentially decaying part. Figure~\ref{Fig:autocorr_calculation}(b) and (c) show the auto-correlation and the fitted model functions of the raw and the corrected auto-correlation data. The underlying frequency of the oscillation can be calculated as the undersampling frequency from the nuclear Larmor frequency $f_\mathrm{L}$ and the sequence length $T_\mathrm{Qd}$ as outlined in \cite{Schmitt2017}.

The measurement sequence, as illustrated in Figure~\ref{Fig:pulse_sequences}(b), consists of optical initialisation, that at the same time acts as readout, and subsequent dynamical decoupling. For the dynamical decoupling a XY8 protocol is embedded in two $\pi/2$-pulses with orthogonal phases. The order of the XY8 block is set between 8 and 24 depending on the depth of the NV centre.

Details to the measurement parameters are shown in \ref{tab:Qdyne_details}. The results of measurement number 1 are presented and discussed in the main text and Figure~4. All other measurements are analysed in the same fashion where the results are shown in Figure~5. For the measurements number 1 to 5 fluorescence-free immersion oil from Fluka is used and for the last one the high-viscosity perfluoropolyether. %The diffusion coefficient of the oils is taken from \cite{Pham2016} to calculate the diffusion time. 

\subsubsection{Additional results}

% Santi and Nicolas

\begin{table}[]
\caption{Experimental parameters for Qdyne measurements.}
\begin{tabular}{ |c|c|c|c|c|c|c| } 
 \hline
 Meas no. & $\delta T_\phi$ & $d$ (nm) & $f_\mathrm{L}$ (kHz) & XY8 order & $T_\mathrm{Qd}$ ($\mu$s) & $T_\mathrm{tot} (h)$ \\ 
 \hline
 1 &\ 0.36 & 15.4 & 2009 & 24 & 49.740 &\ 90 \\  %March 22
 2 &\ 3.00 & 15.4 & 2010 & 22 & 45.607 & 290 \\  %March 26
 3 &\ 4.80 & 15.4 & 2010 & 12 & 25.516 & 170 \\  %April 12
 4 &\ 1.25 &\ 8.0 & 2009 &\ 8 & 17.524 &\ 50 \\  %May 5
 5 &\ 1.59 &\ 8.1 & 2009 &\ 8 & 17.552 &\ 80 \\  %May 12
 6 & 16.70 & 11.4 & 1898 & 12 & 27.788 &\ 65 \\  %June 8
 \hline
\end{tabular}
\label{tab:Qdyne_details}
\end{table}

\begin{figure}[t!]
\includegraphics[width=\columnwidth]{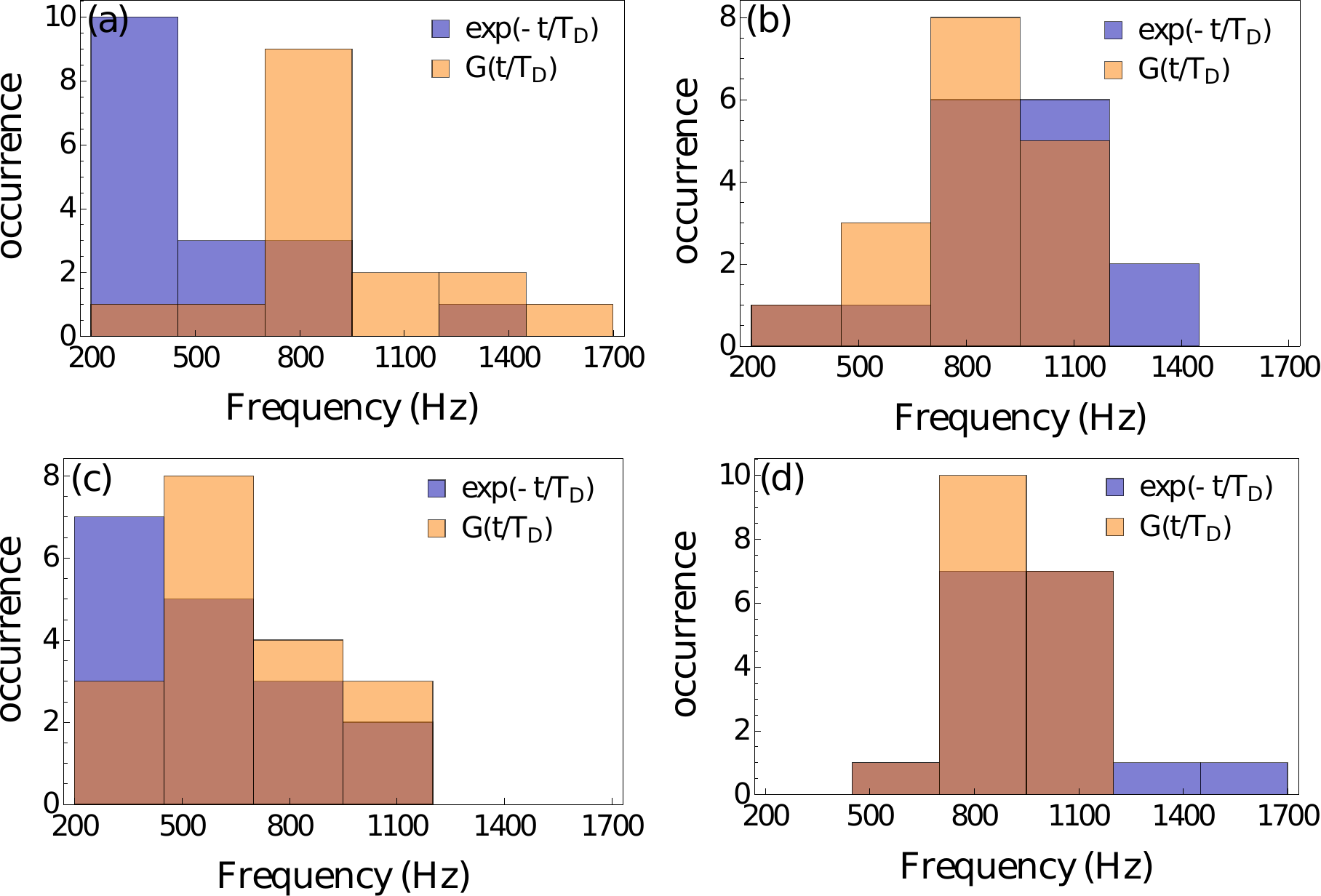}
    \caption{Histograms of frequency estimators from non-linear least squares fitting of 18 auto-correlations from experimental Qdyne time-traces. In orange, fitting to a correlation with power-law decay $C(t/T_D\gg 1) \sim (t/T_D)^{-3/2}$ while in blue the fitting is to a correlation with exponential decay. (a) corresponds to a global optimization with 500 fittings per auto-correlation, with the best result decided by the highest $R^2$ and with $\varphi$ a free fitting parameter. In (b) the phase $\varphi = 0$ is fixed. (c) displays the histogram for local optimization with fixed initial seeding frequency as taken from FFT of the original full time trace, while (d) does the same but with fixed  $\varphi = 0$.
     }\label{reslimitexperimentalappendix}
\end{figure}

Figure~\ref{reslimitexperimentalappendix} shows the complete histogram analysis for experiment one in \ref{tab:Qdyne_details}. Here we include both the global (upper row) and the local (lower row) optimisation results, both without (left column) and with a fixed phase $\varphi$ (right column) for the non-linear least squares fitting. Subfigures (a) and (b) are also presented in Fig.~5 in the main text). As can be seen, only the power-law model does result in meaningful histograms in all instances, with root-mean-square errors ranging from 292\,Hz in the case of global optimisation without fixing the phase Fig.~\ref{reslimitexperimentalappendix}(a) to 151\,Hz for local optimisation with fixed phase in Fig.~\ref{reslimitexperimentalappendix}(d), with a clear trend of diminishing rmse as the fitting model becomes more accurate. In contrast, the exponential correlations model results in wider histograms. The smallest rmse is obtained in Fig~\ref{reslimitexperimentalappendix}(c) with 275\,Hz, where nonetheless the histogram is displaced to the sideband of the allowed search region, meaning there is no resolved frequency. Only for a fixed phase does the exponential correlations model show meaningful histograms, with rmse, respectively 365 and 350\,Hz for global and local optimisations. Note that the flat histogram limit lies at 404\,Hz. 

\begin{figure}[t!]
    \includegraphics[width=\columnwidth]{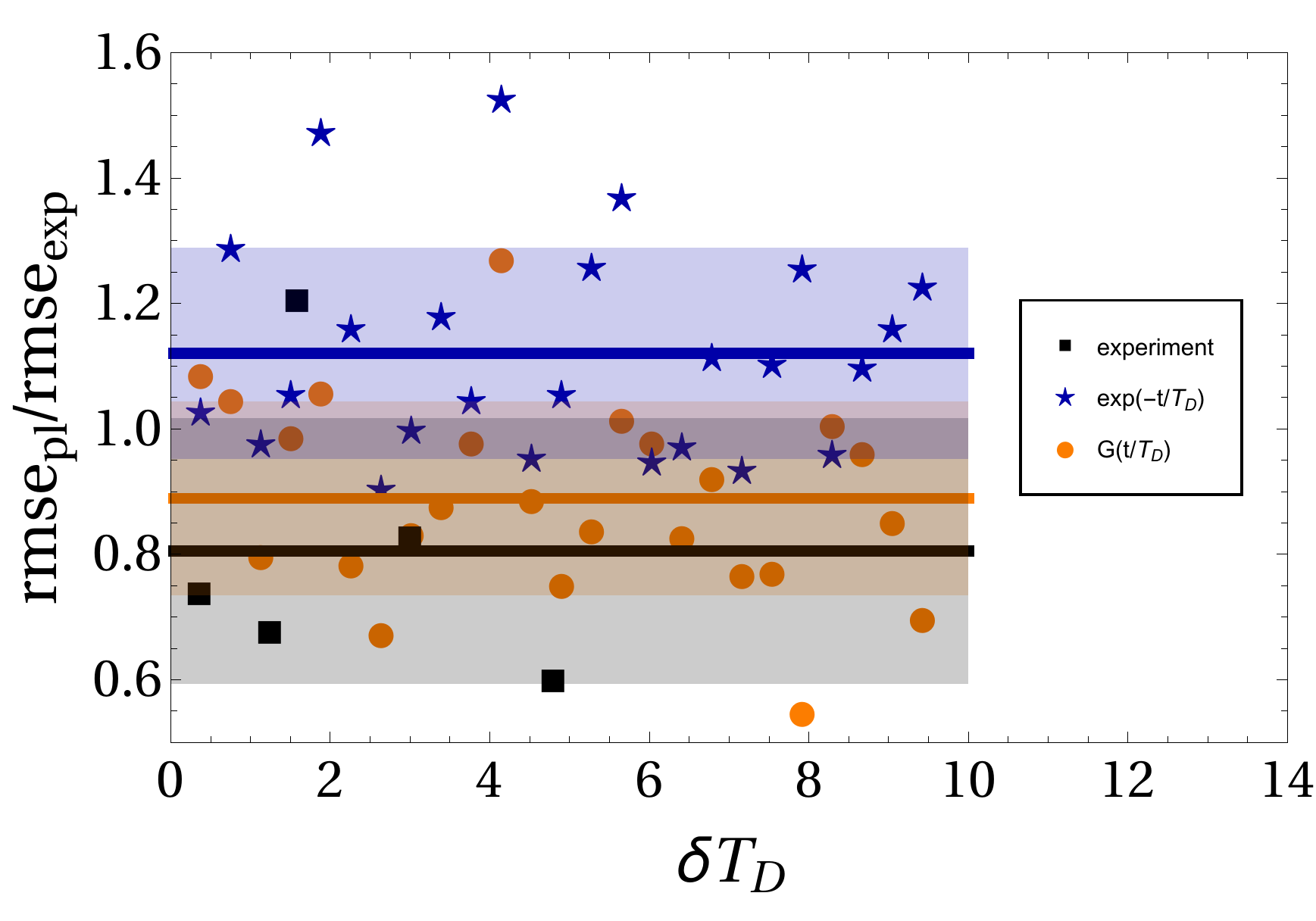}
    \caption{Root-mean-square error ratio between global optimisation for power-law and exponential models fittings. Each dot is calculated with the rmse of histograms as in Fig.~\ref{reslimitexperimentalappendix}. In black squares, experimental results. Orange circles display the ratio for data generated with power-law decay while blue stars correspond to data generated with exponential decay. Solid lines show the mean ratio for all dots while shaded areas correspond to the standard deviation for each mean. Note that experiment six in \ref{tab:Qdyne_details} is not shown but it is considered for the mean ratio.
     }\label{Qdynefigratioglobal}
\end{figure}

In Fig.~\ref{Qdynefigratioglobal} we consider the rmse ratio between \textit{global} optimisation for power-law and exponential fittings, both for all Qdyne experiments as well as numerical simulations of time-traces generated with either of the correlation models and analysed simultaneously with both of them. We observe an experimental mean ratio of 0.81 $\pm$ 0.21, which again favours the interpretation of power-law decay of correlations, since it is this model that better fits the data. Numerical simulations are consistent with this analysis, yielding a ratio of 0.89 $\pm$ 0.16 for correlations generated according to the power-law model while this ratio is 1.12 $\pm$ 0.17 for correlations generated with the exponential model. 

\section{Mixed noise model}\label{Section:Mixed noise model}

Figures 2(b) and 3(b) on the main text show the FFT of the experimental data together with the fits to the exponential and power-law decay models. We see that the latter provides a better fit to the data but nonetheless the experimental FFT has a broadening not directly accounted for by the model. In this section, we hypothesise on the origin of that extra broadening by studying a combined power-law and exponential decay model. 

For a diffusion dominated noise, we expect the main contribution to the correlation function decay to be a power-law, as we have demonstrated in the main text. Other noise sources such as spin diffusion noise by nuclear dipolar interactions are already accounted for by the $T_2$ of the NV centre and, provided measurements do not get close to that time, as happens in our experiments, they do not contribute significantly. Yet there are possible contributions from a magnetic gradient of the NV centre ($\approx 1$ Gauss for correlation spectroscopy with 2.9 nm deep NV and $\approx 0.1$ Gauss for Qdyne), magnetic instabilities, spatial inhomogeneities, or back-action (= $\gamma B N\tau \approx 0.1 < 1$) in the NV centre, that can contribute when measurement times are sufficiently long. These various sources of noise cause exponential decay of the correlation function at a much slower rate than diffusion, and can therefore contribute to a slight Lorentzian broadening of the sharp-peak from the power-law decay. 

\begin{figure}[t!]
    \includegraphics[width=\columnwidth]{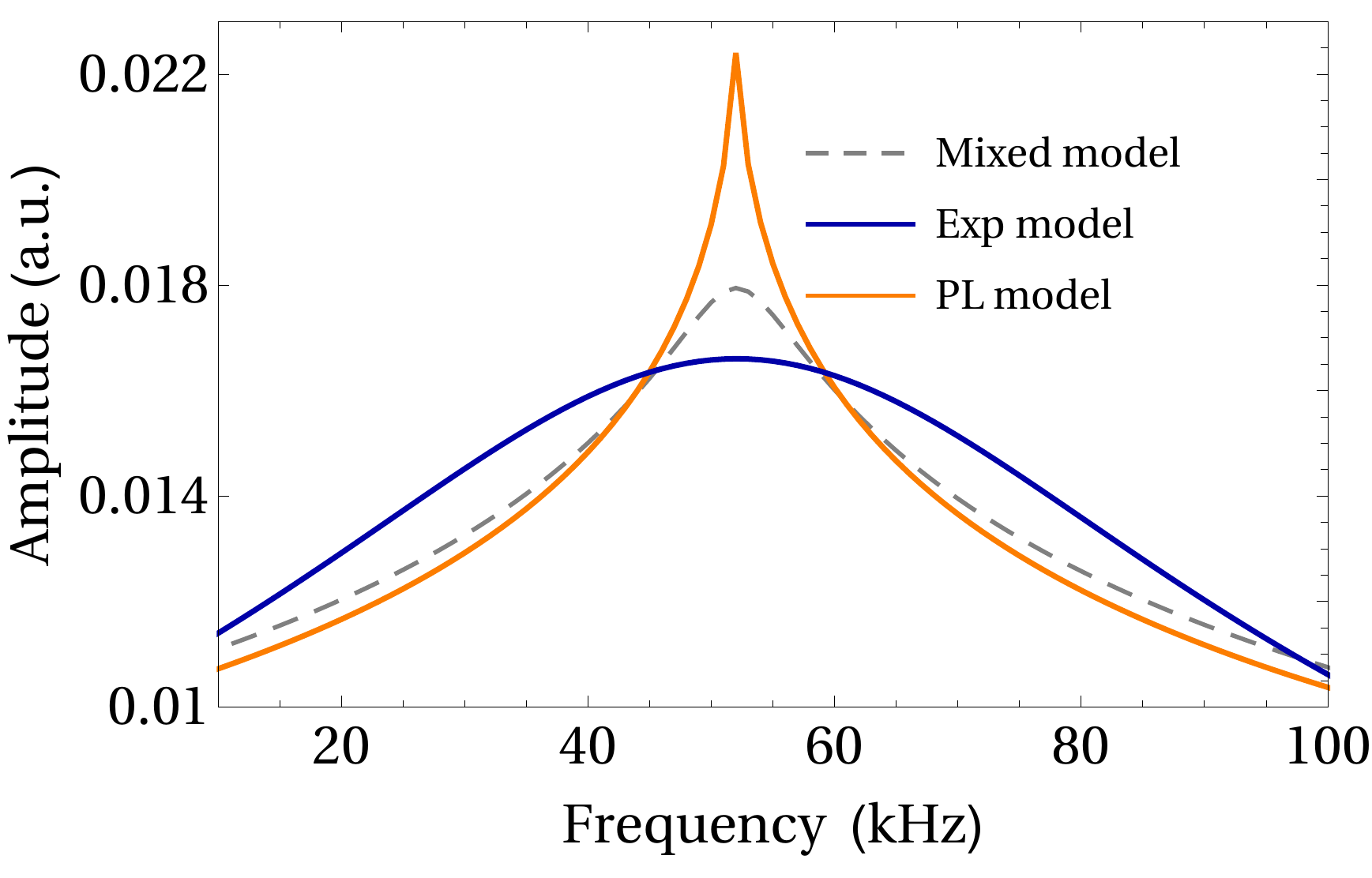}
    \caption{FFT of the correlation function with exponential, power-law, and combined exponential and power-law decays. The first two have a decay rate $T_D$ while for the combined model the power-law decay is characteristic time is $T_D$ and the exponential decay characteristic time is $T_E = 50T_D$.}\label{MixedModelFFT}
\end{figure}

To account for these extra sources, we consider here a model which includes the correlation function decay $G(z)$ from diffusion noise (Eq. (12) in the main text), with an additional exponential decay at a much lower rate, such that $C(z) = G(z)\exp\left(-t/T_E\right)$, with $T_E \gg T_D$. In Fig.~\ref{MixedModelFFT}, we show the FFT for such a mixed model, with $T_E = 50T_D$, together with the power-law decay model and the exponential model both with just with $T_D$ decay rate. We can see that such a mixed model reproduces well the observed extra broadening from the experiments. Note that considering each possible noise source would require studying the particular impact that each of them would have on the correlation function, which need not necessarily be exponential, and estimating their own characteristic time. Such an analysis is however beyond the scope of this work.

\section{Frequency sensitivity for the power-law and exponential models}\label{Section:Frequency sensitivity}

Frequency resolution in a power spectrum is defined as the ability to differentiate two close spectral lines. When a physical process characterised by a frequency is affected by noise, the corresponding spectral line broadens, such that two lines closer than their characteristic linewidth resemble a single, broad spectral line \cite{Gefen2019}. In liquid state nano-NMR this broadening is mainly caused by diffusion. The origin of the spectral line lies in the time-correlated magnetic field generated by a statistically polarised nuclei sample. When such a signal is noisy, correlations decay, limiting the amount of information that can be extracted about the corresponding frequency. We quantify this information by means of the Fisher Information of classical parameter estimation, such that for each specific measurement protocol, and for each model considered for correlations, we have a particular Fisher Information which measures the ability to estimate the frequency, as shown in \cite{Oviedo2020}. 

We can define frequency sensitivity for a specific protocol as the error on the frequency estimation divided by the square root of the total measurement time $\eta = \frac{\Delta \delta}{\sqrt{T_{\text{tot}}}}$. Theoretically, the frequency estimation error is bounded by the inverse square root of the total Fisher Information about the frequency in a given experiment $\Delta\delta \geq \frac{1}{\sqrt{FI_\delta}}$. Thus, considering equal experimental parameters, we can compare the frequency sensitivity in light of each correlations model, i.e. exponential or power-law, through their respective Fisher Informations. 

Considering the ratio of sensitivity in the exponential correlations $\eta_\mathrm{exp}$ to power-law $\eta_\mathrm{pl}$ yields, for correlation spectroscopy $\frac{\eta_\mathrm{exp}}{\eta_\mathrm{pl}} \sim \frac{1}{\delta T_D}$, which diverges for small frequencies $\delta$, meaning that having power-law correlations improves the frequency sensitivity, although globally, for correlation spectroscopy, a resolution problem still exists. For Qdyne, the ratio is $\frac{\eta_\mathrm{exp}}{\eta_\mathrm{pl}} \sim \frac{\log{\left(\delta T_{\text{tot}}\right)}}{\delta^2 T_D^2}$, which not only diverges for small frequencies, but grows with the total measurement time. Thus, power-law correlations significantly improve frequency sensitivity and in principle allow to vanquish the resolution problem \cite{Oviedo2020}. For different experimental protocols, e.g. between correlation spectroscopy and Qdyne, the photon shot noise has an impact on which protocol is optimal, as correlation spectroscopy is less affected. Here as well power-law correlations lessen the impact that such a noise has, as they provide more information for equal total measurement time. 

We can as well estimate the impact on frequency sensitivity from additional noise sources which cause exponential decay at a much slower rate than diffusion. To do so, we calculate the Fisher Information for our combined model in which diffusion induces a power-law decay while other noise sources induce exponential decay of correlations with a characteristic time $T_E \gg T_D$. In this case, for Qdyne measurements \be\nonumber
\frac{\eta_\mathrm{pl}}{\eta_\mathrm{pl+exp}} \sim \frac{\text{Ei}\left(\frac{-T_{\text{tot}}}{T_E}+\delta T_{\text{tot}}\right)+\frac{T_E^3\delta^2\sinh\left(\frac{T_{\text{tot}}}{T_E} \right)}{1+T_E^2\delta^2}}{\log{\left(\delta T_{\text{tot}}\right)}},
\ee
with Ei(x) the exponential integral function, which goes to 0 for small $T_E \rightarrow 0$ and yields $\log{\left(\delta T_{\text{tot}}\right)}$ for large $T_E \rightarrow \infty$.
For frequencies such that $\delta T_D < 1 < \delta T_E$, provided the measurement time is sufficiently long, there is no significant decrease in sensitivity due to these additional noise sources. For frequencies $\delta T_E < 1$ the resolution problem reappears and $\lim_{\delta \rightarrow 0} = 0$. However, note that these additional noise sources contributing to $T_E$, such as, for example, magnetic field instabilities or magnetic defects, contrary to what happens with diffusion, can be mitigated through tweaking the setups, and therefore the resolution problem they lead to is not fundamental in the sense of the diffusion resolution problem which, in the case of exponentially decaying correlations, can only be avoided altering the sample itself, which would forfeit the purpose of nano-NMR. It is expected moreover, that future improvement on setups and experimental techniques leads to the ability to vanquish these noise sources.

\end{document}